\documentclass[prd,aps]{revtex4-1}

\usepackage{amsmath,amssymb,amsfonts,dcolumn,color,graphicx,graphics,latexsym,placeins,epsfig}
\usepackage{subfigure}
\usepackage{natbib}
\newcommand{\be}{\begin{equation}}
\newcommand{\ee}{\end{equation}}
\newcommand{\ba}{\begin{eqnarray}}
\newcommand{\ea}{\end{eqnarray}}

\begin{document}

\title{\Large \bf Born-Infeld gravity coupled to Born-Infeld electrodynamics}

\author{Soumya Jana and Sayan Kar}
\email{soumyajana@phy.iitkgp.ernet.in, sayan@iitkgp.ac.in}
\affiliation{\rm Department of Physics {\it and} Centre for Theoretical Studies \\Indian Institute of Technology, Kharagpur, 721302, India}

\begin{abstract}
\noindent We investigate spherically symmetric, static spacetimes 
in Eddington-inspired Born-Infeld gravity coupled to Born-Infeld 
electrodynamics. 
The two constants, $b^2$ and $\kappa$ which parametrise the Born-Infeld
structures in the electrodynamics (matter) and gravity sectors,
characterise the features of our analytical solutions.
Black holes or naked singularities are found 
to arise, depending on the values of $b^2$ and $\kappa$, 
as well as charge and mass. Several such solutions are classified and
understood through the analysis of the associated metric functions for fixed $\kappa$, varying $b^2$ and vice-versa. Further, we also compare the new metric functions with those for the known $b^2\rightarrow \infty$ (Maxwell) and the $\kappa \rightarrow 0$ (geonic black hole) cases.  
Interestingly, for a particular relation between these
two parameters, $b^2=1/{4\kappa},\, \kappa >0$, we obtain a 
solution resembling the well-known Reissner-Nordstr\" om line element,
albeit some modifications. 
Using this particular solution as the background spacetime, 
we study null geodesics for Born-Infeld photons 
and also, gravitational lensing.
Among interesting features we note $(i)$ an increase in the radius 
of the photon sphere with increasing $\kappa$ and $(ii)$ a net positive contribution in 
the leading order correction term involving $\kappa$, in the weak lensing formula for the deflection angle. We also investigate the effective potential and light propagation for various other solutions through numerics and plots. In summary, our work is the first attempt 
towards figuring out how Born-Infeld
structures in both the matter and gravity sectors can 
influence the nature and character of resulting gravitational fields.      
\end{abstract}


\maketitle

\section{\bf Introduction} 

\noindent General Relativity (GR) is largely 
successful as a classical theory of gravity. It has been tested through 
experiments and observations to very high precision, in vacuum. 
However, it is not entirely certain that GR is the correct theory of gravity 
inside a matter distribution or, in the strong field regime. 
Moreover, it possesses many unsolved puzzles, both theoretical and observational. This has prompted many researchers to actively pursue 
various modified theories of gravity. For example, one motivation for
constructing modified theories of gravity is to look for 
the possibility of avoiding the singularity problem in GR, $eg.$ 
the occurrence of a big bang in cosmology or black holes in an astrophysical
context. In a classical metric theory of gravity, under very general and 
physical assumptions, these singularities are inevitable, as proved through 
the Hawking--Penrose theorems \cite{hawk}. However, it is possible to 
obtain non--singular geometries as solutions in an alternative theory. 
For example, non-singular Friedman--Robertson--Walker (FRW) type cosmologies
do exist in modified gravity theories. 
One can also argue that the singularity in the classical theory 
is expected to be resolved in quantum gravity. Unfortunately, 
though specific examples exist in the context of loop quantum cosmology, 
we are still far away from any general statements 
which may be possible when we have a fully 
consistent quantum description of gravity. 

\noindent In this article, we investigate one such alternate theory 
which is inspired from Eddington's alternative formulation of GR. 
In this formulation, the $\sqrt{-g}R$ (Ricci scalar) in the Einstein-Hilbert
action is replaced by $\sqrt{-det (R_{ij})}$ and the connection is chosen as 
the basic variable, instead of the metric. Thus, Eddington's formulation 
is essentially an affine formulation \cite{edd}. In vacuum, it is equivalent to GR and it describes de Sitter gravity. However, coupling of matter
was never addressed in this formulation until very recently. 

\noindent We are also aware of Born--Infeld electrodynamics \cite{born}
wherein the divergence in the electric field at the location of the charge/current is removed by using an action different from that of Maxwell. 
The new field theory due to Born and Infeld includes the 
determinantal structures in the action, inspired by Eddington's work. 
The virtue of this nonlinear electrodynamics in removing the
singularity of the electric field, inspired Deser and Gibbons 
to construct a Born-Infeld gravity theory in the metric formulation  \cite{desgib}. Later, Vollick \cite{vollick} introduced the Palatini formulation in  
Born-Infeld gravity and worked on various related aspects. 
He also introduced a non-trivial and somewhat artificial way of coupling matter in such a theory \cite{vollick2,vollick3}. More recently, Banados and Ferreira  \cite{banados} have come up with a formulation where the matter coupling is 
different and simpler compared to Vollick's proposal. 
We will focus here on the theory proposed in \cite{banados} and refer to it
as Eddington-inspired Born--Infeld (EiBI) gravity, for obvious reasons. 
Note that the EiBI theory has the feature that it reduces to GR, in vacuum.

\noindent Interestingly, EiBI theory also falls within the class of bimetric 
theories of gravity (also called bi-gravity). The  current bimetric theories 
have their origin in the seminal work of Isham, Salam and Strathdee \cite{isham}. Several articles have appeared in the last few years on various aspects of 
such bi-gravity theories. In \cite{scargill}, the authors have pointed out that the EiBI field equations can also be derived from an equivalent bi-gravity action. This action is closely related to a recently discovered family of unitary 
massive gravity theories which are built as bi-gravity theories. 
Several others  have contributed in this direction, in various ways 
\cite{schmidt,lavinia}.   

\noindent Let us now briefly recall Eddington--inspired Born--Infeld gravity. 
The central feature here is the existence of a physical metric which couples to matter and another auxiliary metric which is not used for matter couplings. 
One needs to solve for both metrics through the field equations. The action for the theory developed in \cite{banados}, is given as:
\begin{equation}
S_{BI}(g,\Gamma, \Psi) =\frac{1}{8\pi\kappa}\int d^4 x \left [ \sqrt{-\vert g_{\mu\nu} +\kappa R_{\mu\nu}(\Gamma)\vert}-\lambda \sqrt{-g} \right]+ S_M (g, \Psi)
\end{equation}
where $\Lambda= \frac{\lambda-1}{\kappa}$. We assume $G=1$ and $c=1$ throughout in the paper. Variation w.r.t $\Gamma$, done using the
auxiliary metric $q_{\mu\nu}=g_{\mu\nu}+\kappa R_{\mu\nu}(q)$ gives
\begin{equation}
q_{\mu \nu}=g_{\mu\nu} + \kappa R_{\mu\nu}(q)
\label{eq:gammavarn}
\end{equation}
Variation w.r.t $g_{\mu\nu}$ gives 
\begin{equation}
\sqrt{-q} q^{\mu\nu} = \lambda \sqrt{-g}g^{\mu\nu}-8\pi\kappa \sqrt{-g} T^{\mu\nu}
\label{eq:gvarn}
\end{equation}
In order to obtain solutions, we need to assume a $g_{\mu\nu}$ and 
a $q_{\mu\nu}$
with unknown functions, as well as a matter stress energy ($T^{\mu\nu}$). 
Thereafter, we write down the field equations and obtain solutions using 
some additional assumptions about the metric functions and the stress energy. 

\noindent Quite some work on various fronts have been carried out
on various aspects of this theory, in the last few years.
Astrophysical scenarios have been discussed  in 
\cite{cardoso,casanellas,avelino,sham,sham2,structure.exotic.star,sotani.neutron.star,
sotani.stellar.oscillations,sotani.magnetic.star} while
cosmology has been dealt with in  \cite{escamilla,cho,avelinoferreira,felice,scargill,linear.perturbation,power.spectrum,large.scale.structure,
bianchi.cosmo}. Among other topics,
a domain wall brane in a higher dimensional generalisation, 
has been analysed in \cite{eibibrane}. 
Generic features of the paradigm of matter-gravity couplings
have been analysed in \cite{delsate}. In \cite{cho_prd88}, authors have shown that EiBI theory admits some nongravitating matter distribution, which is not allowed in GR.  Some interesting cosmological and 
circularly symmetric solutions in $2+1$ dimensions are shown in \cite{jana}. 
Constraints on the EiBI parameter $\kappa$ are obtained  
from studies on compact stars \cite{cardoso,sham2}, tests in the 
solar system \cite{solar.test}, astrophysical and cosmological observations \cite{avelino}, and nuclear physics studies \cite{nuclear.test}. 
In \cite{pani}, a major problem related to surface singularities has been 
noticed in the context of  stellar physics. However, gravitational 
backreaction has been suggested as a cure to this problem 
in \cite{eibiprob.cure}. In \cite{odintsov}, the authors propose a modification to EiBI theory by taking its functional extension in a way similar to $f(R)$ theory. Recently, the authors in \cite{fernandes} use a different way of matter coupling, by taking the Kaluza ansatz for five dimensional EiBI action in a purely metric formulation and then compactify it using Kaluza's procedure to get a four dimensional gravity coupled in a nonlinear way to the electromagnetic theory. 

\noindent 
Not much has however been done on black hole physics, or, broadly on 
the spherically symmetric, static solutions in this theory. It may be noted 
that the vacuum, spherically symmetric static solution is trivially same as 
the Schwarzschild de Sitter black hole. But, the electrovacuum solutions are 
expected to deviate from the usual Reissner-Nordstr\"om solution in GR. 
This has been shown in \cite{banados,wei,sotani} where the authors 
consider EiBI gravity coupled to a Maxwell electric field of a 
localized charge. They obtain the resulting spacetime geometries, and study its properties. The basic features of such spacetimes includes a singularity 
at the location of the charge which may or may not be covered by an 
event horizon. The strength of the electric field remains 
nonsingular as in Born-Infeld electrodynamics. 
However, this may not be the only solution because, in EiBI gravity, 
the matter coupling is nonlinear. The authors of \cite{eibiwormhole} have 
shown, in a different framework, that the central singularity could be 
replaced by a wormhole supported by the electric field. In \cite{rajibul}, the author obtained a class of Lorentzian regular wormhole spacetimes supported by the quintessential matter which does not violate the weak or null energy cindition in EiBI gravity.

\noindent In our present work, we couple EiBI gravity with 
Born-Infeld electrodynamics. Thus, we have Born-Infeld structures in both
the gravity and matter sectors. 
We obtain some new classes of spherically symmetric static spacetimes and study their properties. The solutions we present include black holes and naked 
singularities. Earlier, a lot of work has indeed been done by considering 
nonlinear electrodynamics coupled to GR \cite{geon,Einstein.BI.1,Einstein.BI.2,Einstein.BI.3,dibakar1,dibakar2,dibakar3,jonas}. Some of them are motivated by string theory since 
Born-Infeld structures naturally arise in the low energy limit of 
open string theory \cite{openstring1,openstring2}. Our article 
is sectioned as follows. In Section II, we derive the form of the 
stress-energy tensor assuming an ansatz for the physical metric ($g_{\mu\nu}$). 
We obtain the general expressions for the metric functions 
by solving the EiBI field equations with an ansatz for the 
auxiliary metric ($q_{\mu\nu}$). In Section III, we list the solutions 
by scanning across regions in the parameter space ($b^2,\, \kappa$), 
where $b^2$ and $\kappa$ are the parameters in the electromagnetic and gravity
sectors, respectively. In Section IV, we analyze, in detail, 
a particular class of solutions of the Reissner-Nordstr\"om-type 
for which $b^2=\frac{1}{4\kappa}$ ($\kappa >0$). Here we also discuss null geodesics, photon propagation, and gravitational 
lensing. Finally, in Section V, we investigate the effective potential and gravitational lensing for the $b^2\neq \frac{1}{4\kappa} $ solutions. In Section VI, we summarize our results and indicate
possibilities for future work.                

\section{Spherically symmetric static spacetimes due to a charged mass}
\label{sec:cosmo}
In curved spacetime, the Lagrangian density for the Born-Infeld electromagnetic field theory is given by \cite{born},
\begin{equation}
\mathcal{L}=\frac{b^2\sqrt{-g}}{4\pi}\left[1-\sqrt{1+\frac{F}{b^2}-\frac{\mathcal{G}^2}{b^4}}\right]
\end{equation} 
where, $F=\frac{1}{2}F_{\mu\nu}F^{\mu\nu}$ and 
$\mathcal{G}=\frac{1}{4}F_{\mu\nu}\mathcal{G}^{\mu\nu}$ are two scalar 
quantities constructed from the components of the electromagnetic field 
tensor ($F_{\mu\nu}$) and the dual field tensor ($\mathcal{G}_{\mu\nu}$). 
Here, $b$ sets an upper limit on the electromagnetic field and, 
when $b\rightarrow \infty$,  Maxwell's theory is recovered. 
The resulting energy-momentum tensor has the following general expression:
\begin{equation}
T_{\mu\nu}=-\frac{2}{\sqrt{-g}}\frac{\partial \mathcal{L}}{\partial g^{\mu\nu}}=-\frac{b^2}{4\pi}\left[g_{\mu\nu}\left(\sqrt{1+\frac{F}{b^2}-\frac{\mathcal{G}^2}{b^4}}-1\right)-\frac{b^2F_{\mu\sigma}F^{\, \sigma}_{\nu}-\mathcal{G}^2g_{\mu\nu}}{b^4\sqrt{1+\frac{F}{b^2}-\frac{\mathcal{G}^2}{b^4}}}\right]
\end{equation}  
In our work here we restrict ourselves to an  
electrostatic scenario (i.e. $A_{\mu}\equiv \lbrace\phi,0,0,0\rbrace$, $\mathcal{G}=0$), 
in the background of a spherically symmetric, static spacetime. We 
assume the line element to be of the form,
\begin{equation}
ds^2=-U(\bar{r})e^{2\psi(\bar{r})}dt^2+U(\bar{r})e^{2\nu(\bar{r})}d\bar{r}^2+V(\bar{r})\bar{r}^2\left(d\theta^2+\sin^2
\theta d\phi^2\right)
\label{eq:physical_metric_ansatz}
\end{equation}
where, $\bar{r}$ is the radial coordinate, but the physical radial distance or, the area radius of each 2-sphere is $r=\bar{r}\sqrt{V(\bar{r})}$. 
With this assumption, solving the equations of motion for the electromagnetic 
field, we obtain
\begin{equation} 
\frac{d\phi}{d\bar{r}}=-\frac{qUe^{\nu+\psi}}{\sqrt{V^2\bar{r}^4+q^2/b^2}}
\label{eq:dphi_dr} 
\end{equation}
were, $q$ is the electric charge.  In the Minkowski limit 
of Eq.~(\ref{eq:physical_metric_ansatz}), the Eq.~(\ref{eq:dphi_dr}) 
resembles the known Born-Infeld electric field due to a point charge. 
The non-zero components of the energy-momentum tensor are given as:
\begin{eqnarray}
T_{tt}&=&\frac{b^2U}{4\pi}\left(\frac{\sqrt{V^2\bar{r}^4+q^2/b^2}}{V\bar{r}^2}-1 \right)e^{2\psi} , \nonumber \\
T_{\bar{r}\bar{r}}&=&-\frac{b^2U}{4\pi}\left(\frac{\sqrt{V^2\bar{r}^4+q^2/b^2}}{V\bar{r}^2}-1 \right)e^{2\nu},\nonumber \\
T_{\theta\theta}&=&\frac{b^2}{4\pi}\left(1-\frac{V\bar{r}^2}{\sqrt{V^2\bar{r}^4+q^2/b^2}} \right)V\bar{r}^2,\nonumber \\
\mbox{and,}\quad T_{\phi\phi}&=&\frac{b^2}{4\pi}\left(1-\frac{V\bar{r}^2}{\sqrt{V^2\bar{r}^4+q^2/b^2}} \right)V\bar{r}^2\sin^2\theta.
\label{eq:energymomentum_tensor}
\end{eqnarray}
The energy-momentum tensor is not traceless ( $i.e.$ $g^{\mu\nu}T_{\mu\nu}\neq 0 $ for any $q/b\neq 0$). 
The auxiliary line element is assumed to be of the form
\begin{equation}
ds^2_q=-e^{2\psi}dt^2+e^{2\nu}d\bar{r}^2+\bar{r}^2\left(d\theta^2+\sin^2\theta d\phi^2\right)
\label{eq:auxiliary_metric_ansatz}
\end{equation}
Here, the area radius of the auxiliary spacetime is $\bar{r}$.
It is worth noting that our choice of the physical and auxiliary line 
elements are opposite to the standard choices in \cite{banados,wei,sotani,eibiwormhole}. While making this choice, we have exploited two facts: $(i)$ the absence of differential expressions in the field equations in~(\ref{eq:gvarn}) 
which opens up the scope of algebraic manipulations, and $(ii)$ 
the specific forms of the $tt$- and $\bar{r}\bar{r}$- components of the stress-energy 
tensor. This choice simplifies the situation to quite an extent. 
Using Eqs.~(\ref{eq:physical_metric_ansatz}),(\ref{eq:energymomentum_tensor}), 
and (\ref{eq:auxiliary_metric_ansatz}) in the field equations obtained 
from the `$g$'-variation [Eq.~(\ref{eq:gvarn})], and after some algebra 
, we obtain the expressions for $U$ and $V$. We get a quadratic equation for $V$ as,
\begin{equation}
\left(1-4\kappa b^2\right)V^2-2\left(1-2\kappa b^2\right)V+1-\frac{4\kappa^2 q^2 b^2}{\bar{r}^4}=0
\label{eq:v_qudra}
\end{equation}
There are two possible solutions for $V$ from Eq.~(\ref{eq:v_qudra}) and consequently, two solutions for $U$. 
We choose the solution for $U$ and $V$ such that $U\rightarrow 1, \,V\rightarrow 1$ for large $\bar{r}$ and the physical spacetime is that found in the GR regime. Therefore, we have,
\begin{eqnarray}
V(\bar{r})&=&\frac{1-2\kappa b^2\left(1+\sqrt{1+\frac{q^2}{b^2\bar{r}^4}-\frac{4\kappa q^2}{\bar{r}^4}} \right)}{1-4\kappa b^2}\label{eq:Veqn} \\
U(\bar{r})&=&\frac{(1-2\kappa b^2)\sqrt{1+\frac{q^2}{b^2\bar{r}^4}-\frac{4\kappa q^2}{\bar{r}^4}}-2\kappa b^2}{(1-4\kappa b^2 )\sqrt{1+\frac{q^2}{b^2\bar{r}^4}-\frac{4\kappa q^2}{\bar{r}^4}}} \label{eq:Ueqn}
\end{eqnarray} 
From the field equations obtained via `$\Gamma$'-variation [Eq.~(\ref{eq:gammavarn})], we get,
\begin{eqnarray}
\kappa e^{-2\psi}R_{tt}(q)-U=-1,\label{eq:R_tt_qmetric}\\
\kappa e^{-2\nu}R_{\bar{r}\bar{r}}(q)+U=1, \label{eq:R_rr_qmetric} \\
\mbox{and}\quad~ V\bar{r}^2+\kappa \left[-\frac{d}{d\bar{r}}\left(\bar{r}e^{-2\nu} \right)+1 \right]&=&\bar{r}^2 ,
\label{eq:R_thth_qmetric}
\end{eqnarray}  
where, $R_{tt}(q)=e^{-2\nu+2\psi}\left(\frac{2\psi'}{\bar{r}}-\nu'\psi'+\psi'^2+\psi'' \right)$ and $R_{\bar{r}\bar{r}}(q)=\frac{2\nu'}{\bar{r}}+\nu'\psi'-\psi'^2-\psi''$. Here, 
primes denote derivatives with respect to $\bar{r}$. From Eq.~(\ref{eq:R_tt_qmetric}) and Eq.~(\ref{eq:R_rr_qmetric}), we get $\psi'+\nu'=0$. So, without loss of any generality, we can assume $\nu=-\psi$. Thus, from the Eq.~(\ref{eq:R_thth_qmetric}), we arrive at,
\begin{equation}
e^{2\psi(\bar{r})}=1-\frac{\bar{r}^2}{3\kappa}+\frac{1}{\kappa \bar{r}}\int V(\bar{r})\bar{r}^2d\bar{r}+\frac{C}{\bar{r}}
\label{eq:e2psi}
\end{equation}
where, $C$ is a constant of integration. For, $q=0$ ($i.e.$ vacuum), $U=V=1$ [Eqs.~(\ref{eq:Ueqn}),(\ref{eq:Veqn})] for any $\kappa$, and the solution should be the
Schwarzschild spacetime. Therefore, $C$ is identified as related to the 
total mass, $i.e.$, $C=-2M$.

\section{Classification of spacetimes}
\label{sec:classification}
\subsection{General Solutions}
In this section, we obtain different classes of solutions by scanning 
the $b^2-\kappa$ parameter space. 
For this purpose, we introduce the definition $b^2=\frac{\alpha}{4\kappa}$. 
We note 
that, if $\kappa>0 $, then $\alpha>0$, and if $\kappa <0$ then $\alpha <0$. The characteristics/forms of the solutions change over different ranges of 
$\alpha$. These are--($i$) $\alpha\rightarrow \infty$, ($ii$) $\infty > \alpha >1 $, ($iii$) $\alpha \rightarrow 1$, ($iv$) $-\infty <\alpha < 1$ (but $\neq 0$), and ($v$) $\alpha \rightarrow -\infty$. We discuss the $\alpha =1$ case elaborately, as a special case, in a separate section. The remaining 
classes of solutions are analysed here. 
Firstly, we note that $\alpha \rightarrow \pm \infty$ cases are actually those
belonging to 
the Maxwell limit, $i.e.$, a coupling of EiBI gravity with a 
Maxwellian electric field. This has already been studied earlier \cite{wei,sotani,eibiwormhole}. One can verify that the solution for this case 
(particularly for $\alpha \rightarrow \infty$), after rewriting in the 
Schwarzschild gauge, takes exactly the same form as that of \cite{wei}. 
Next, we obtain the solutions for the cases ($ii$) and ($iv$) below:\\

First, we rewrite $U$, $V$ as,
\begin{eqnarray}
U(\bar{r})&=&\frac{2-\alpha}{2(1-\alpha)}-\frac{\alpha}{2(1-\alpha)}\frac{1}{\sqrt{1+\frac{4\kappa q^2(1-\alpha)}{\alpha \bar{r}^4}}},\label{eq:U_alpha}\\
V(\bar{r})&=& \frac{2-\alpha}{2(1-\alpha)}-\frac{\alpha}{2(1-\alpha)}\sqrt{1+\frac{4\kappa q^2(1-\alpha)}{\alpha \bar{r}^4}}.\label{eq:V_alpha}
\end{eqnarray}   
Using, the Eq.~(\ref{eq:V_alpha}), we compute the integral $\left(\int V\bar{r}^2d\bar{r}\right)$ (see the Appendix) and use it in the Eq.~(\ref{eq:e2psi}) to find the expression of $e^{2\psi(\bar{r})}$.

\

{\em{($ii$)$\infty > \alpha >1 $ case}}: 

Using Eq.~(\ref{eq:vint1}), we get,
\begin{eqnarray}
e^{2\psi(\bar{r})}&=&1+\frac{\alpha \bar{r}^2}{6\kappa (\alpha -1)}\left[\sqrt{1-\frac{4\kappa q^2(\alpha -1)}{\alpha \bar{r}^4}}-1\right]+ 
\frac{\alpha^{1/4}(4q^2)^{3/4}}{3\kappa^{1/4}(\alpha -1)^{1/4}\bar{r}}F\left(\arcsin\left(\frac{\left(4\kappa q^2(\alpha -1)\right)^{1/4}}{\alpha^{1/4}\bar{r}}\right) \middle \vert -1\right) \nonumber\\
&&-\frac{2M}{\bar{r}}
\label{eq:e2psi_classii}
\end{eqnarray}
where, $F(\phi\vert m)=\int^{\phi}_0[1-m\sin^2\theta]^{-1/2}d\theta$ is the incomplete elliptic integral of the first kind.
Thus we find all the metric functions. Here, the radius-squared of the 
2-sphere at each value of the radial coordinate $\bar{r}$ is given by $r^2(\bar{r})=V(\bar{r})\bar{r}^2$. We note that, for $1<\alpha\leq 2$, at $\bar{r}=(\alpha \kappa q^2)^{1/4}$, the radius-square ($r^2$) becomes zero. But, for $\alpha >2$, there is a non-zero minimum value of $r^2$, $i.e.$ $r^2(\bar{r}_0)=\frac{(\alpha -2)\bar{r}_0^2}{2(\alpha -1)}$, where $\bar{r}_0=\left[4\kappa q^2 (1-1/\alpha)\right]^{1/4}$. This means, for $1<\alpha \leq 2$, the spacetime is due to a point charge ($q$) of total mass $M$, but, for $\alpha >2$, it is distributed over a spherical shell in space. One can verify that, 
for a very small value of $\kappa$, the solution converges to the known 
Reissner-Nordstr\"om solution. The reason for this is that, as $\kappa$ becomes 
very small with respect to the value of $\alpha$, $b^2$ becomes large, which then 
is the Maxwell limit for the electric field and the GR limit in the 
gravity sector. Further, we have examined the singularities of the spacetime 
and found that curvature scalar ($\mathcal{R}$) diverges at the {\em position} 
of the charge. Moreover, the existence of the horizon depends upon the values 
of the parameters $\alpha, \kappa, q$, and $M$. Some of these features are demonstrated in Fig.~\ref{fig:compare_maxwell1} and Fig.~\ref{fig:compare_geon}, 
where the plots of the metric functions in the Schwarzschild coordinates are 
shown.

\noindent In the Schwarzschild gauge, the line element [eq.~(\ref{eq:physical_metric_ansatz})] becomes
\begin{eqnarray}
ds^2&=&-g_{tt}dt^2+g_{rr}dr^2+r^2\left(d\theta^2 + \sin ^2\theta d\phi^2 \right) \nonumber \\
\mbox{where,}\quad~ g_{tt}&=&U(\bar{r})e^{2\psi(\bar{r})},\quad~ g_{rr}=\frac{V(\bar{r})}{U(\bar{r})}e^{-2\psi(\bar{r})},\,
\mbox{and,}\quad~ \bar{r}=r\left[1-\frac{\alpha}{2}+\frac{\alpha}{2}\sqrt{1+\frac{4\kappa q^2}{\alpha r^4}} \right]^{1/2} 
\label{eq:schw_gauge}
\end{eqnarray}  

{\em ($iv$)$-\infty<\alpha<1$ case }:

In this case, using Eq.~(\ref{eq:vint2_hyper}) we get,

\begin{eqnarray}
e^{2\psi (\bar{r})}&=&1-\frac{2M}{\bar{r}}-\frac{\alpha \bar{r}^2}{6\kappa (1-\alpha)}\left[\sqrt{1+\frac{4\kappa q^2 (1-\alpha)}{\alpha \bar{r}^4}}-1 \right]+ \frac{4q^2}{3\bar{r}^2}\, {}_{2}F_1\left(\frac{1}{4},\frac{1}{2};\frac{5}{4};-\frac{4\kappa q^2 (1-\alpha)}{\alpha \bar{r}^4}\right)
\label{eq:e2psi_classiv}
\end{eqnarray} 
where, ${}_{2}F_1(a,b;c;z)$ is the hypergeometric function with usual meaning. 
Here, for $0<\alpha<1$, the $r^2(\bar{r}_0)=0$, for $\bar{r}_0=(\alpha \kappa q^2)^{1/4}$. But, for $\alpha<0$ ($i.e.$ for $\kappa <0$), there is a minimum value of the radius-square, which is $r^2(\bar{r}_0)=\left(\frac{\kappa q^2 \alpha}{1-\alpha}\right)^{1/2}$, for $\bar{r}_0=0$. Therefore, for $\alpha <0$ or $\kappa <0$, the charge ($q$) and the mass ($M$) are distributed over a 2-sphere, but, for $1>\alpha>0$, it is a point charge. This solution also reduces to the Reissner-Nordstr\"om solution for very small $\kappa$ ($\kappa \rightarrow \pm 0$). Here also, the spacetime is singular at the location of the charge and it may be a black hole or naked singularity depending upon the parameter values. The metric functions for this class of solutions are plotted in the Schwarzschild coordinates [eq.~(\ref{eq:schw_gauge})] for some parameter values and these are shown in Fig.~\ref{fig:compare_maxwell2} and Fig.~\ref{fig:compare_geon}.

\subsection{Deviation from the solution with Maxwell field}
For $\kappa>0$, the solution with the Maxwell field ($\alpha\rightarrow \infty$) is given by (using Eqs.~[\ref{eq:U_alpha},\ref{eq:V_alpha},\ref{eq:e2psi_classii},\ref{eq:schw_gauge}]),
\begin{eqnarray}
ds^2&=&-\frac{r^4h_1(r)}{(r^4-\kappa q^2)}dt^2+\frac{(r^4-\kappa q^2)}{(r^4+\kappa q^2)h_1(r)}dr^2+r^2\left(d\theta^2+\sin^2\theta d\phi^2\right)\label{eq:metric_kpos_maxwell} \\
\mbox{where},\, h_1(r)&=& 1-\frac{q^2}{3r^2}-\frac{2Mr}{\sqrt{r^4+\kappa q^2}}+\frac{r(4q^2)^{3/4}}{3\kappa^{1/4}\sqrt{r^4+\kappa q^2}}F\left(\arcsin\left(\frac{(4\kappa q^2 r^4)^{1/4}}{\sqrt{r^4+\kappa q^2}} \right)\middle\vert -1\right). \label{eq:h1}
\end{eqnarray} 
Since, the charge is distributed over a 2-sphere of radius $r_0=(\kappa q^2)^{1/4}$, therefore $r\geq (\kappa q^2)^{1/4}$. For $\kappa <0$, the solution with the Maxwell field ($\alpha \rightarrow -\infty$) is given by the Eq.~(\ref{eq:metric_kpos_maxwell}), but with the replacement of $h_1(r)$ by $h_2(r)$, where
\begin{equation}
h_2(r)= 1-\frac{q^2}{3r^2}-\frac{2Mr}{\sqrt{r^4+\kappa q^2}}+ \frac{4q^2r^2}{3(r^4+\kappa q^2)}\,\, {}_{2}F_{1}\left(\frac{1}{4}, \frac{1}{2}; \frac{5}{4}; \frac{4\kappa q^2 r^4}{(r^4+\kappa q^2)^2}\right) \label{eq:h2}
\end{equation} 

\begin{figure}[!htbp]
\centering
\subfigure[$\kappa=1.0, q=0.5, M=1.0$]{\includegraphics[width=3.0in,angle=360]{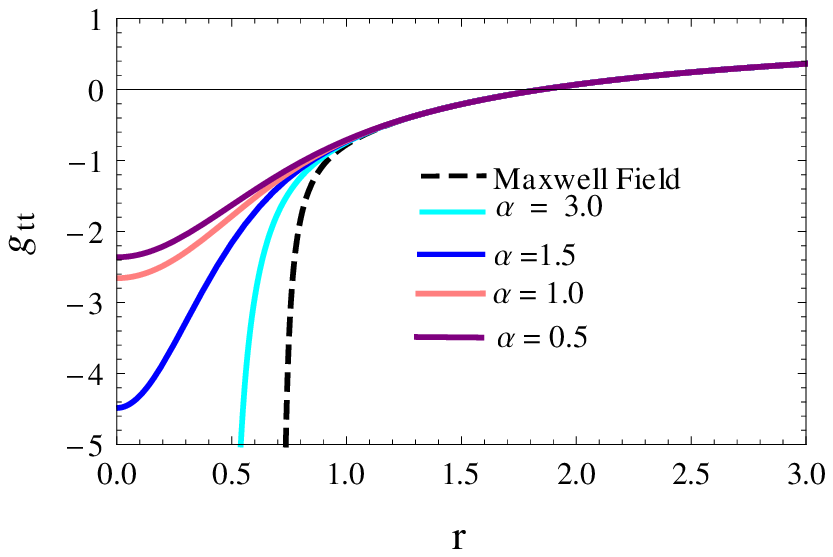}\label{subfig:gtt1}}
\subfigure[$\kappa=1.0, q=0.5, M=1.0$]{\includegraphics[width=3.0in,angle=360]{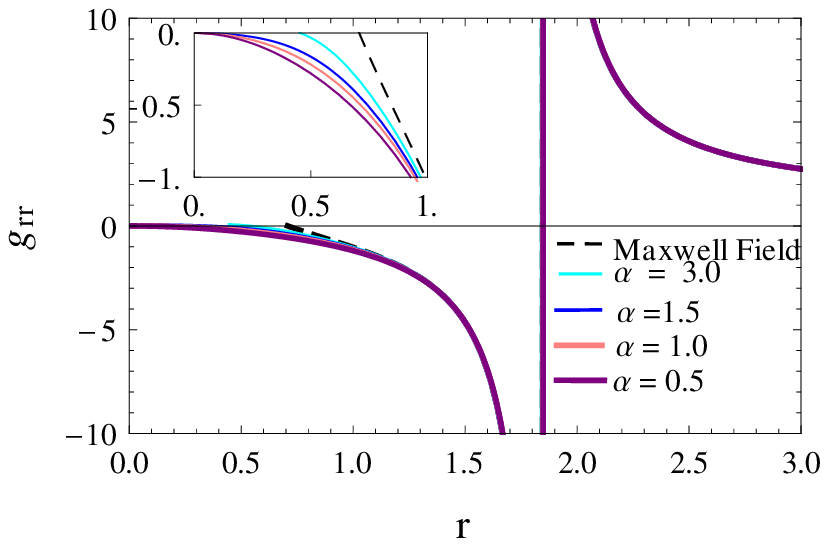}\label{subfig:grr1}}
\subfigure[$\kappa=1.0, q=0.9, M=1.0$]{\includegraphics[width=3.0in,angle=360]{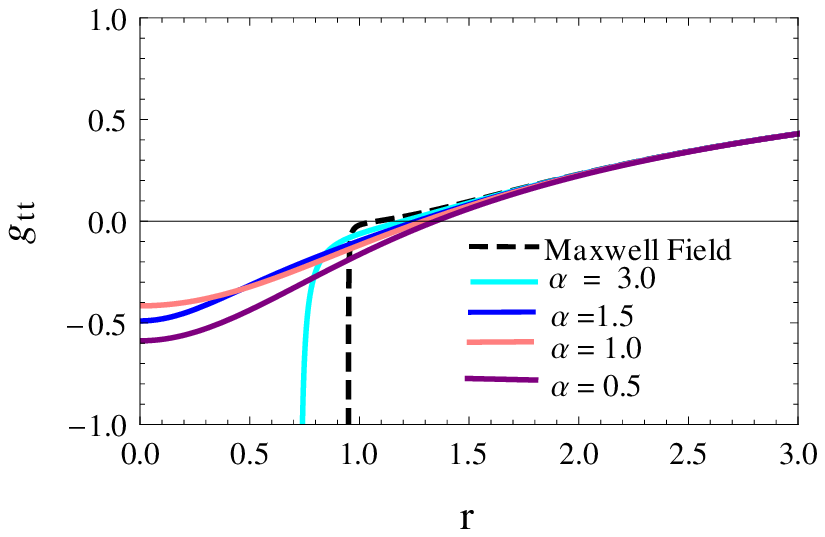}\label{subfig:gtt2}}
\subfigure[$\kappa=1.0, q=0.9, M=1.0$]{\includegraphics[width=3.0in,angle=360]{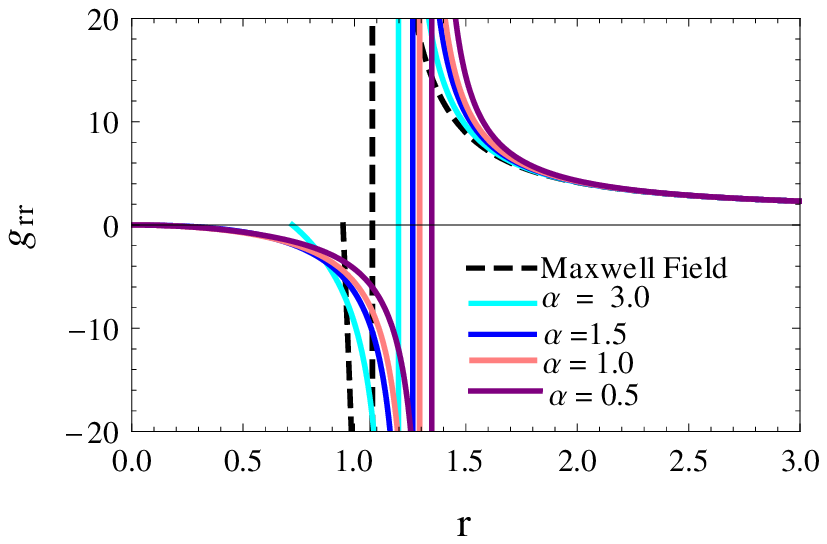}\label{subfig:grr2}}
\caption{Metric functions in Schwarzschild coordinates are plotted and compared with the EiBI solution for the Maxwell electric field (black dashed line). $g_{tt}$ and $g_{rr}$ are the functions given in the equation (\ref{eq:schw_gauge}). In the plots, $\kappa=1.0$, while $\alpha=4\kappa b^2$ takes different values, which means the solutions are with different values of control parameter ($b^2$) for the nonlinear Born-Infeld electric field.}
\label{fig:compare_maxwell1}
\end{figure}

\begin{figure}[!htbp]
\centering
\subfigure[$\kappa=-1.0, q=0.5, M=1.0$]{\includegraphics[width=3.0in,angle=360]{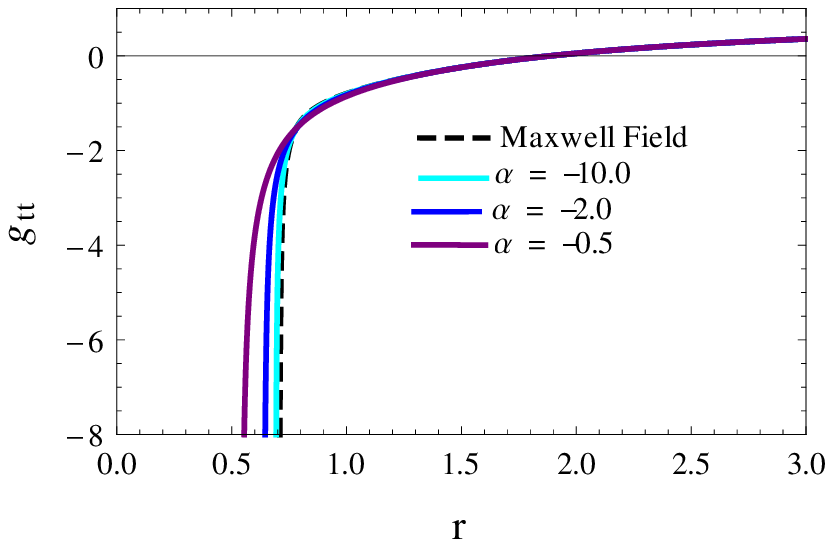}\label{subfig:gtt3}}
\subfigure[$\kappa=-1.0, q=0.5, M=1.0$]{\includegraphics[width=3.0in,angle=360]{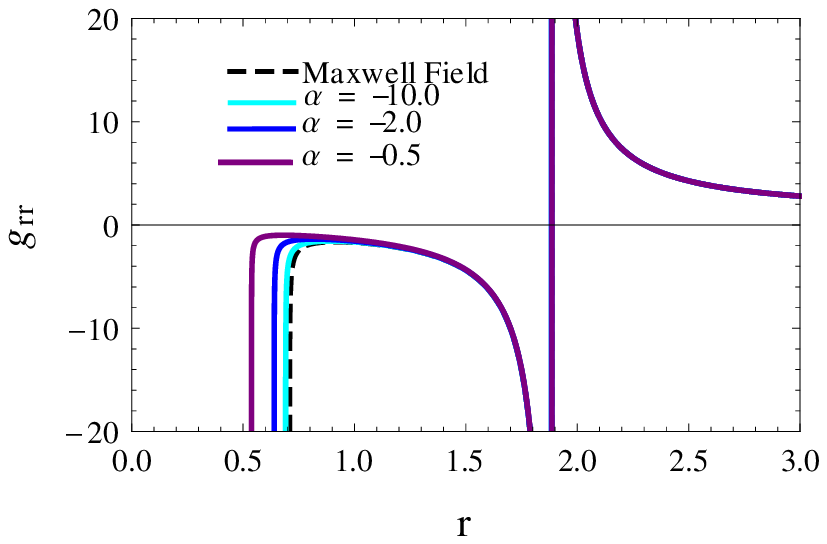}\label{subfig:grr3}}
\subfigure[$\kappa=-1.0, q=0.9, M=1.0$]{\includegraphics[width=3.0in,angle=360]{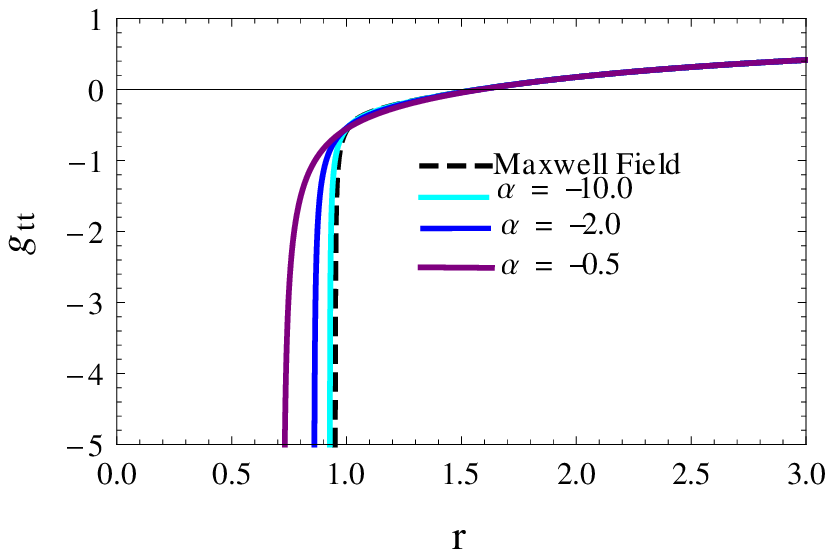}\label{subfig:gtt4}}
\subfigure[$\kappa=-1.0, q=0.9, M=1.0$]{\includegraphics[width=3.0in,angle=360]{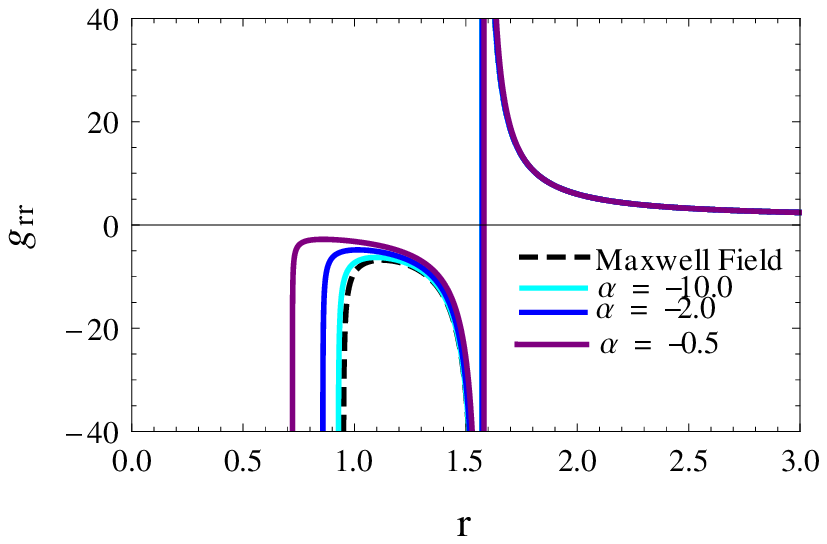}\label{subfig:grr4}}
\caption{Metric functions in Schwarzschild coordinates are plotted and compared with the EiBI solution for the Maxwell electric field (black dashed line). $g_{tt}$ and $g_{rr}$ are the functions given in the equation (\ref{eq:schw_gauge}). Here, in these plots also, $\kappa$-value is fixed but negative and consequently $\alpha$ is also negative.  }
\label{fig:compare_maxwell2}
\end{figure}

The deviation of the EiBI solutions with Born-Infeld electric field from that with the Maxwell field are demonstrated in Fig.~\ref{fig:compare_maxwell1} and Fig.~\ref{fig:compare_maxwell2} for $\kappa$ positive and negative respectively. In each figure, $\kappa$-value is fixed, but $\alpha=4\kappa b^2$ takes different values for different solid lines. Therefore, Born-Infeld electric field parameter ($b^2$) is varied without disturbing the gravity sector.  All plots show the variation of metric functions in the Schwarzschild coordinates with physical radial distance $r$. The plots are obtained using Eqs.~[\ref{eq:U_alpha}, \ref{eq:V_alpha}, \ref{eq:e2psi_classii},\ref{eq:e2psi_classiv}, \ref{eq:schw_gauge}, \ref{eq:phyline_alpha1_schwgauge}, \ref{eq:metric_kpos_maxwell}, \ref{eq:h1}, \ref{eq:h2}]. For a given value of $\alpha$, an appropriate aforementioned equation has been used. In some plots $r$ starts with a nonzero minimum value which is due to the charges distributed over a 2-sphere of finite radius.

\noindent In Fig.~\ref{subfig:gtt1}, we clearly see that the metric function ($g_{tt}$) for different values of $\alpha= 4\kappa b^2$ deviate from the case with a Maxwell field (black dashed line) at small $r$. However, at large $r$, plots for different $\alpha$ approach the black dashed line (Maxwell) coinciding with it asymptotically. We also see that, for larger values of $\alpha$ or $b^2$, the amount of deviation reduces. In Fig.~\ref{subfig:gtt1} and Fig.~\ref{subfig:grr1}, we note that for the chosen parameter values ($\kappa=1.0, q=0.5, M=1.0$) the deviation in the horizon radius from the case with a Maxwell field is negligible. We note a clear deviation ( Fig.~\ref{subfig:gtt2} and Fig.~\ref{subfig:grr2}) in the horizon radius for a different value of $ q=0.9$. Here, the horizon radius increases with decreasing $\alpha$, or $b^2$.  In Fig.~\ref{fig:compare_maxwell1}, one can see that for $\alpha=0.5$ and $\alpha=1.5$ the metric functions (both $g_{tt}$ and $g_{rr}$) begin from $r=0$ indicating a point charge. In contrast, for $\alpha=3.0$ and the Maxwell limit (black dashed line), the metric functions are limited to a minimum radius of the 2-sphere where the charge is distributed. From the Fig.~\ref{fig:compare_maxwell2}, one can also note the existence of a minimum radial distance for all $\alpha$ with  $\kappa$ negative.        

\subsection{Deviation from the Geonic blackhole solution}
Additionally, we note that, in the limits-- $\alpha\rightarrow \pm 0$ and $\kappa \rightarrow \pm 0$, but for a finite $b^2$ we get back the geonic black hole solution. The geonic blackhole is a solution in GR coupled to a 
Born-Infeld electric field due to a point charge \cite{geon}. In a geonic black hole scenario, a distant observer associates a total mass which comprises 
$M$ (the black hole mass) and a pure electromagnetic mass stored as the 
self energy in the electromagnetic field. If $M$ is zero, the spacetime becomes regular everywhere. 

\begin{figure}[!htbp]
\centering
\subfigure[$b^2=1.0, q=0.9, M=1.0$]{\includegraphics[width=3.0in,angle=360]{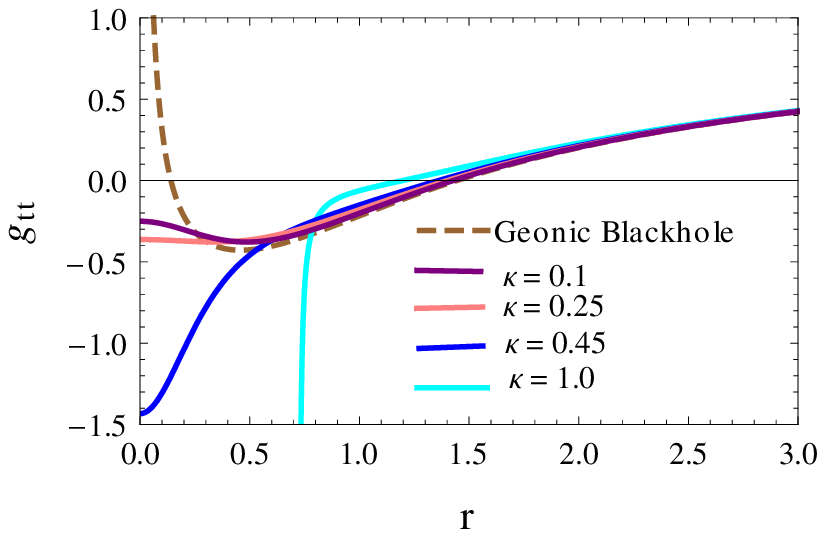}\label{subfig:geongtt1}}
\subfigure[$b^2=1.0, q=0.9, M=1.0$]{\includegraphics[width=3.0in,angle=360]{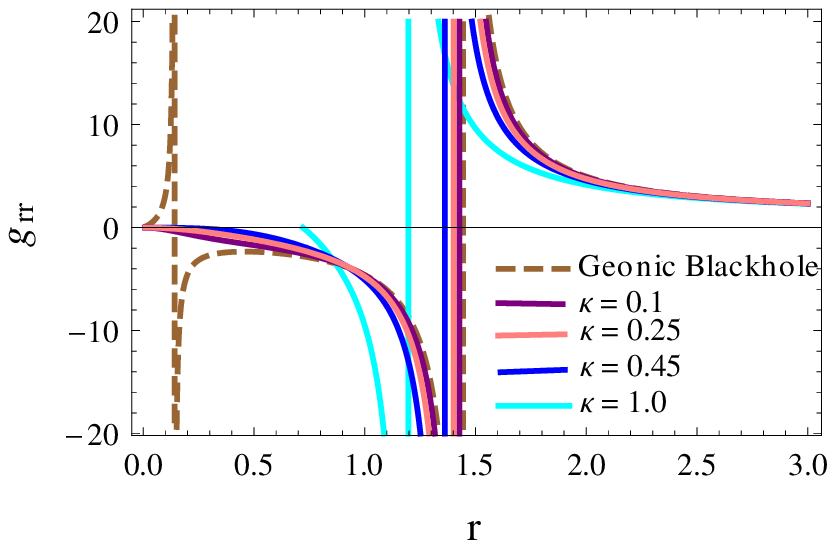}\label{subfig:geongrr1}}
\subfigure[$b^2=1.0, q=0.5, M=1.0$]{\includegraphics[width=3.0in,angle=360]{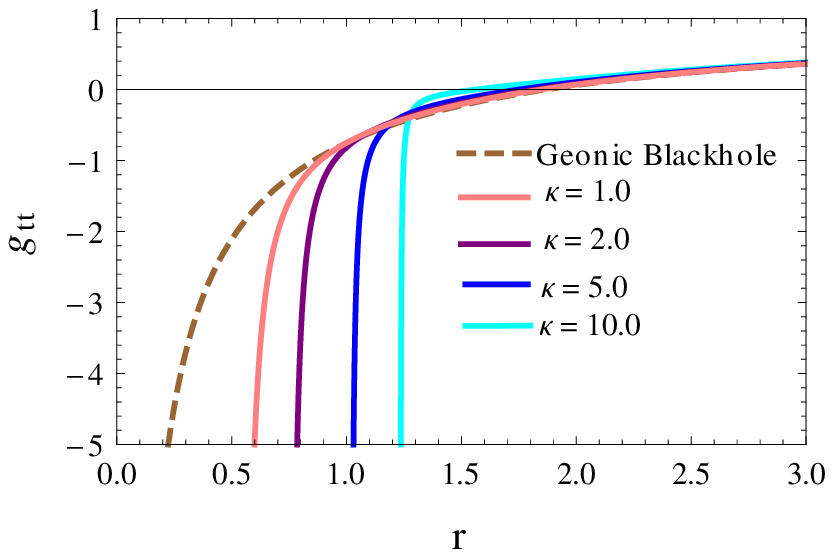}\label{subfig:geongtt2}}
\subfigure[$b^2=1.0, q=0.5, M=1.0$]{\includegraphics[width=3.0in,angle=360]{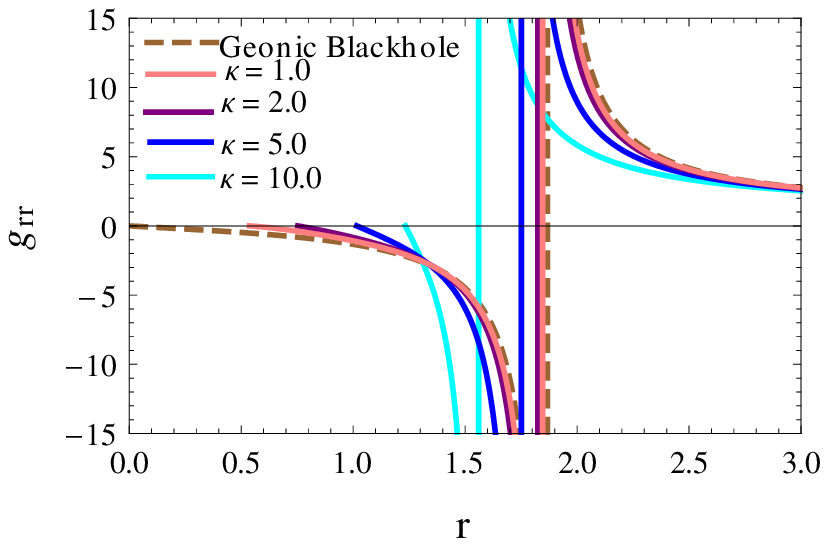}\label{subfig:geongrr2}}
\subfigure[$b^2=1.0, q=0.7, M=1.0$]{\includegraphics[width=3.0in,angle=360]{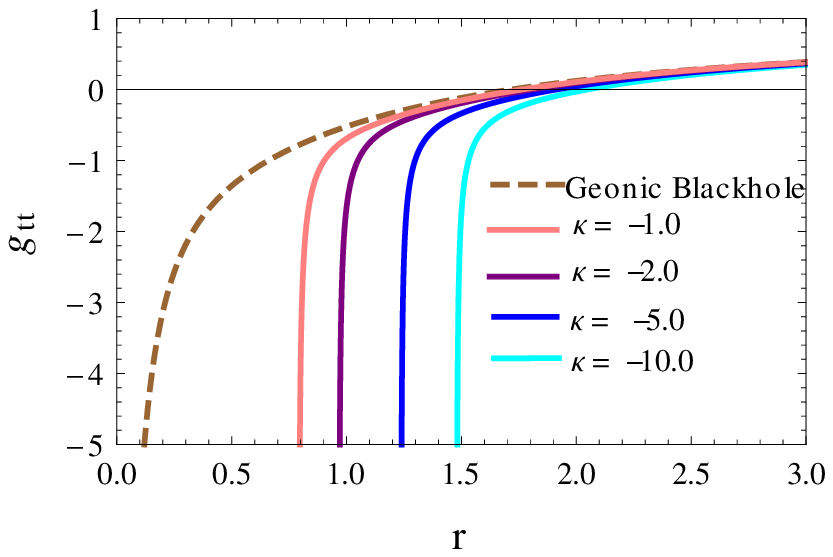}\label{subfig:geongtt2}}
\subfigure[$b^2=1.0, q=0.7, M=1.0$]{\includegraphics[width=3.0in,angle=360]{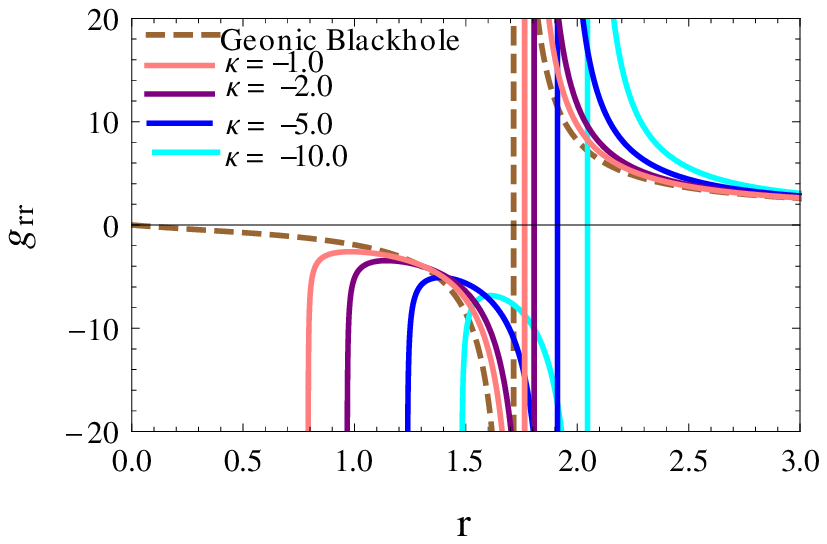}\label{subfig:geongrr3}}
\caption{Metric functions in Schwarzschild coordinates are plotted and compared with the geonic blackhole solution (brown dashed line). $g_{tt}$ and $g_{rr}$ are the functions given in the equation (\ref{eq:schw_gauge}). Here, in these plots, the Born-Infeld electric field parameter ($b^2$) has the fixed value $b^2=1$, but EiBI gravity parameter $\kappa$ is varied.}
\label{fig:compare_geon}
\end{figure}

The line element for a geonic blackhole is given as (from Eq.~\ref{eq:e2psi_classiv}),
\begin{eqnarray}
ds^2&=&-g_e(r)dt^2+\frac{dr^2}{g_e(r)}+r^2(d\theta^2 +\sin^2\theta d\phi^2)
\label{eq:metric_geon}\\
\mbox{where,}\,\, g_e(r)&=& 1-\frac{2M}{r}-\frac{2q^2}{3\left(\sqrt{r^4+q^2/b^2}+r^2\right)}+\frac{4q^2}{3r^2}\, {}_{2}F_1\left(\frac{1}{4}, \frac{1}{2}; \frac{5}{4}; -\frac{ q^2}{b^2r^4}\right) \label{eq:ge}
\end{eqnarray} 
where, $r$ is again the physical radial distance. Here, $r$ can take any positive value from zero to infinity. The deviation of the EiBI solutions (with different $\kappa$) from the geonic blackhole solution (the GR limit) is demonstrated in Fig.~\ref{fig:compare_geon} with a fixed value of the Born-Infeld electric field parameter ($b^2$). Here, the gravity sector is varied by changing the value of $\kappa$ without disturbing the matter sector. 

The deviation in the metric functions from the geonic blackhole solution (brown dashed line, $\kappa=0$) is prominent for different values of $\kappa$ in Fig.~\ref{fig:compare_geon}. However, at larger $r$, all plots approach the brown dashed line. For a larger magnitude of $\kappa$, the deviation is more. In Fig.~\ref{subfig:geongtt1} and Fig.~\ref{subfig:geongrr1}, we see that for the given parameter values, the double horizon exists in the case of the geonic blackhole, though it is absent for other solutions with different values of $\kappa$. From Fig.~\ref{subfig:geongrr1}, \ref{subfig:geongrr2}, and \ref{subfig:geongrr3}, we note a generic feature that for $\kappa$ positive the horizon radius decreases with increase in magnitude of $\kappa$. But, it is the opposite for negative $\kappa$.

\section{The $\alpha=1$ case}
\subsection{The line element}
\noindent For $\alpha = 1$, the quadratic equation [Eq.~(\ref{eq:v_qudra})] becomes linear in $V$. Hence, the metric functions $V$, $U$ and $e^{2\psi}$ [Eqs.~(\ref{eq:U_alpha}),(\ref{eq:V_alpha}), (\ref{eq:e2psi})] become
\begin{equation}
 V=1-\frac{\kappa q^2}{\bar{r}^4}, \quad~ U=1+\frac{\kappa q^2}{\bar{r}^4},\quad~ \mbox{and}\quad~ e^{2\psi}=1-\frac{2M}{\bar{r}}+\frac{q^2}{\bar{r}^2}
\label{eq:metricfuncn_alpha1}
\end{equation}
Then, the physical line element becomes,
\begin{equation}
ds^2=-\left(1+\frac{\kappa q^2}{\bar{r}^4} \right)\left( 1-\frac{2M}{\bar{r}}+\frac{q^2}{\bar{r}^2} \right)dt^2+\frac{\left(1+\frac{\kappa q^2}{\bar{r}^4} \right)}{\left( 1-\frac{2M}{\bar{r}}+\frac{q^2}{\bar{r}^2} \right)}d\bar{r}^2+\left(1-\frac{\kappa q^2}{\bar{r}^4} \right)\bar{r}^2\left(d\theta^2+\sin^2\theta d\phi^2\right)
\label{eq:phyline_alpha1}
\end{equation}
Here, the radius-square: $r^2(\bar{r})=\bar{r}^2- \frac{\kappa q^2}{\bar{r}^2}$, becomes zero at $\bar{r}=(\kappa q^2)^{1/4}$. So, this describes a point charge $q$ with the total mass $M$. This line element [Eq.~(\ref{eq:phyline_alpha1})] clearly shows that, for $\kappa =0$, it becomes the Reissner-Nordstr\"om solution. The auxiliary line element is also the same as Reissner-Nordstr\"om. 

The source for such a spacetime has the following form of the stress-energy tensor in the framework of EiBI gravity:
\begin{equation}
T^{\mu}_{\; \nu}=diag.\left\lbrace -\frac{q^2}{8\pi(\bar{r}^4-\kappa q^2)}, -\frac{q^2}{8\pi(\bar{r}^4-\kappa q^2)}, \frac{q^2}{8\pi(\bar{r}^4+\kappa q^2)},\frac{q^2}{8\pi(\bar{r}^4+\kappa q^2)}\right\rbrace.
\label{eq:energymomentum_alpha1}
\end{equation}
One may verify that the stress energy satisfies the weak/null energy conditions \cite{wald}.

\begin{figure}[!htbp]
\centering
\subfigure[]{\includegraphics[width=3.25in,angle=360]{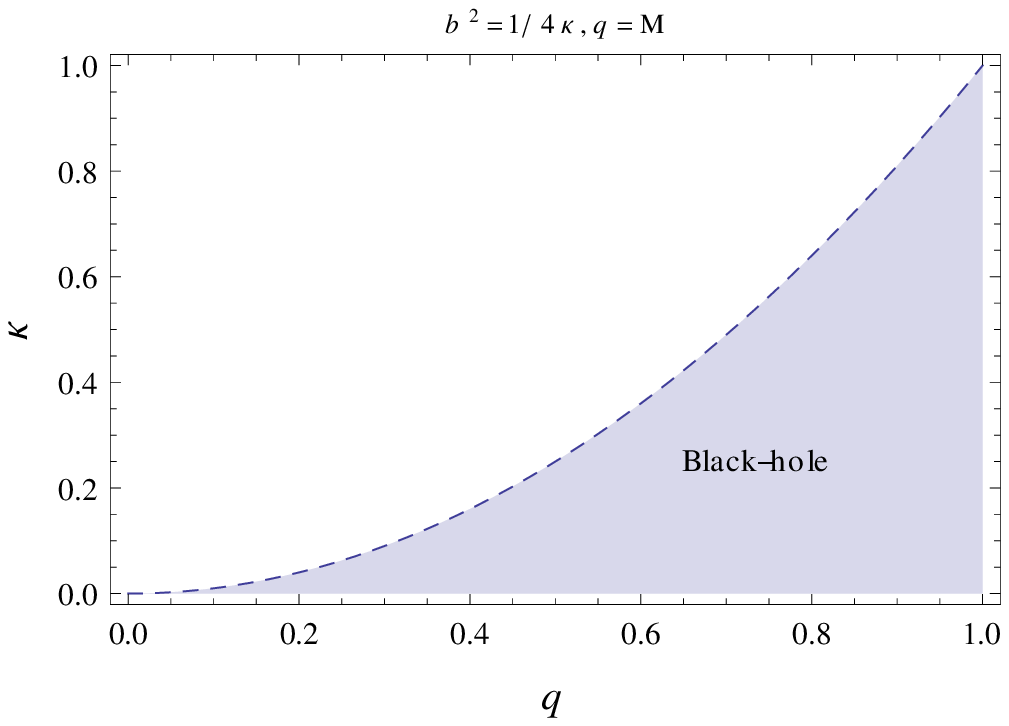}\label{subfig:horizon_alpha11}}
\subfigure[]{\includegraphics[width=3.40in,angle=360]{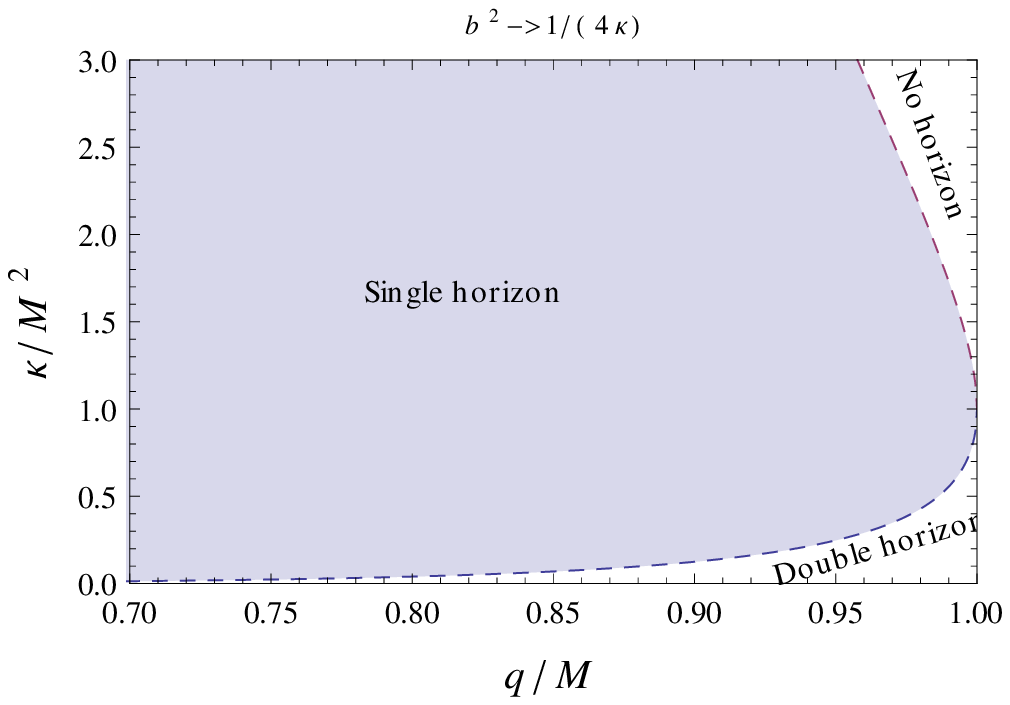}\label{subfig:horizon_alpha12}}

\caption{(Left panel)the parameter space: $q^2-\kappa$ showing the existence/non-existence of horizon, for $q=M$. (Right panel) the same plot for $q\neq M$ case.}
\label{fig:hz_alpha1}
\end{figure}

In the Schwarzschild gauge, the line element takes the form
\begin{equation}
ds^2=-f(r)dt^2+\frac{2r^2}{\left(r^2+\sqrt{r^4+4\kappa q^2} \right)f(r)}dr^2+r^2\left(d\theta^2 +\sin^2 \theta d\phi^2 \right)
\label{eq:phyline_alpha1_schwgauge}
\end{equation} 
where, the redshift function, $f(r)=\left(1+\frac{4\kappa q^2}{\left(r^2+\sqrt{r^4+4\kappa q^2} \right)^2}\right)\left[1-\frac{2\sqrt{2}M}{\sqrt{r^2+\sqrt{r^4+4\kappa q^2}}}+\frac{2q^2}{r^2+\sqrt{r^4+4\kappa q^2}} \right]$. At the horizon-radius, $r_{hz}$, $f(r_{hz})=0$. The expression for the horizon-radius is therefore given by,
\begin{eqnarray}
r^2_{hz}=\frac{\left( M \pm \sqrt{M^2-q^2}\right)^4-\kappa q^2}{\left(M\pm \sqrt{M^2-q^2}\right)^2}
\label{eq:hz_alpha1}
\end{eqnarray}
Here, we note that a horizon does not exist for $q>M$, similar to the Reissner-Nordstr\"om case. For $q=M$, we have $r^2_{hz}=q^2-\kappa$. So, unlike the Reissner-Nordstr\"om extremal solution, for each value of $q=M$, there is an upper limit on $\kappa$ for the existence of the horizon, (see the left panel of the Fig.~\ref{fig:hz_alpha1}). From the right panel of the Fig.~\ref{fig:hz_alpha1}, we note that, for $q<M$, for sufficiently small values of $\kappa$ there may be double horizon. For a range of higher values of $\kappa$ a single-horizon exists, and for even more high values of $\kappa$ the 
horizon vanishes. However, in the Reissner-Nordstr\"om counterpart, we always 
have the double horizon.  


\begin{figure}[!htbp]
\centering
\subfigure[]{\includegraphics[width=3.0in,angle=360]{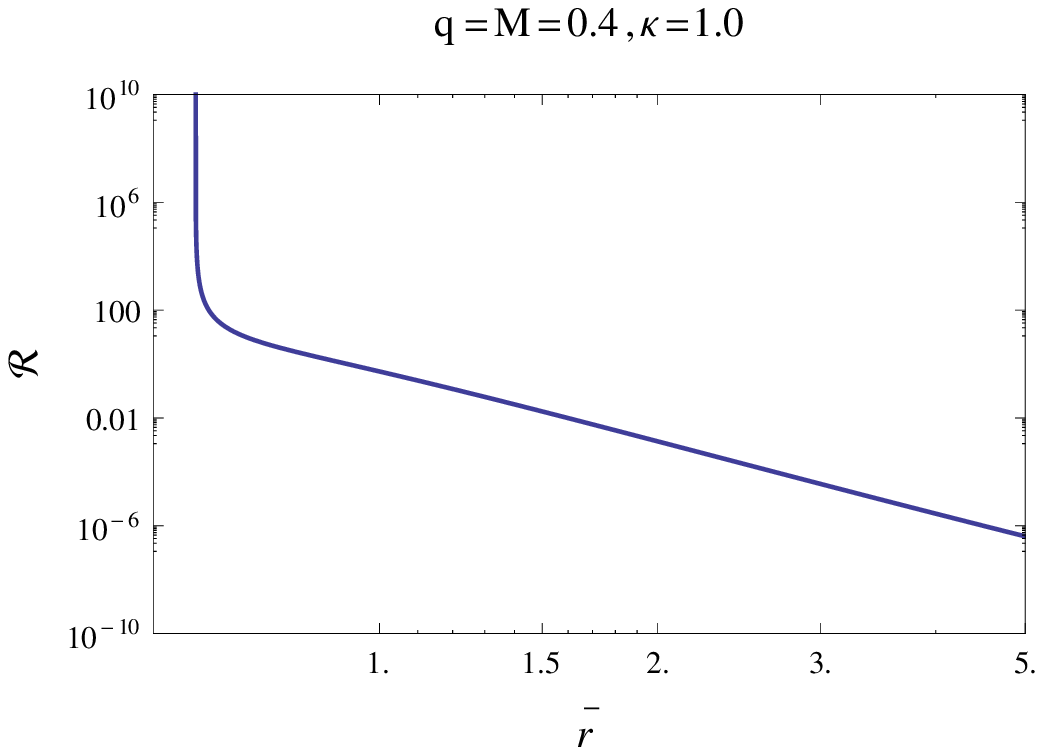}\label{subfig:singluarity_alpha11}}
\subfigure[]{\includegraphics[width=3.0in,angle=360]{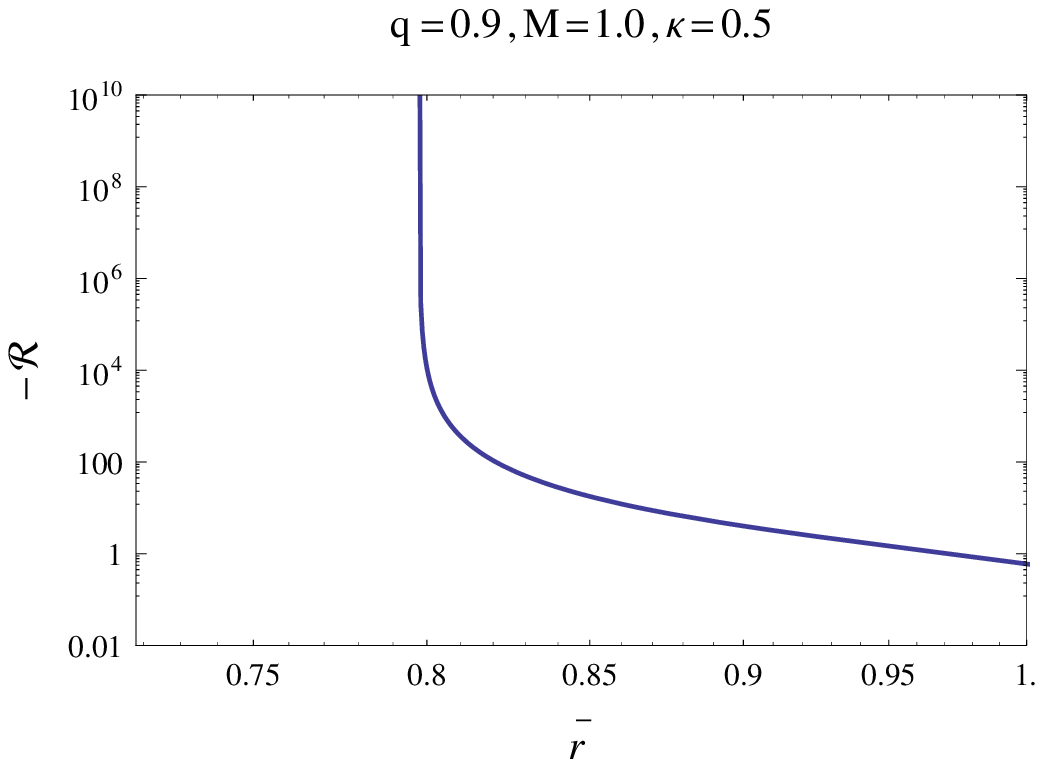}\label{subfig:singluarity_alpha12}}

\caption{(Left panel)the ricci scalar ($\mathcal{R}$) is plotted as a function of $\bar{r}$ for the parameter values, $q=M=0.4,\, \kappa=1.0$. The plot shows that $\mathcal{R}$ diverges at the location of the charge, $i.e.\, \bar{r}=(\kappa q^2)^{1/4}\, or,\, r=0$. (Right panel) the same plot for a $q\neq M$ case. Here, $\mathcal{R}$ diverges to negative values.}
\label{fig:singluarity_alpha1}
\end{figure}

One can verify that, in general, all the scalar quantities, like the Ricci scalar 
($\mathcal{R}=g^{ij}R_{ij}$), $R_{ij}R^{ij}$, and the Kretschmann scalar 
($R_{ijkl}R^{ijkl}$), diverge at the location of the point charge, 
$i.e.$ at $\bar{r}=(\kappa q^2)^{1/4}$, or, $r=0$. However, for a special choice, $q=\sqrt{\kappa}/3$ and $M=2\sqrt{\kappa}/{3\sqrt{3}}$, the Ricci scalar becomes regular everywhere. But, the spacetime still remains singular at the location of the point charge, for such a choice, since the Kretschmann scalar and $R_{ij}R^{ij}$  diverge. Fig.~\ref{fig:singluarity_alpha1} 
demonstrates the singularity for two different value of the parameters. So, the spacetime 
is eventually singular, but may or may not be covered by a horizon. However, 
we show that the strength of the electric field at the location of the 
charge remains finite. For this, we consider a static observer having the four 
velocity, $u^{\mu}=\lbrace e^{-\psi}/\sqrt{U}, 0,0,0 \rbrace$. The electric field vector is defined as $E_{\mu}=u^{\nu}F_{\mu\nu}$. So, the square of the 
strength of the electric field is 
$E^2=E^{\mu}E_{\mu}=\frac{q^2}{V^2\bar{r}^4+q^2/b^2}$. At the location of the charge, $E^2=\frac{q^2}{r^4(\bar{r}_0)+q^2/b^2}=b^2=1/{4\kappa}$ (in this case, $r(\bar{r}_0)=0$). 
The energy density observed by such a static observer is $\rho_{obs}=u^{\mu}u^{\nu}T_{\mu\nu}=\frac{b^2}{4\pi}\left(\frac{\sqrt{V^2\bar{r}^4+q^2/b^2}}{V\bar{r}^2}-1 \right)$. In this case, $\rho_{obs}=\frac{q^2}{8\pi(\bar{r}^4-\kappa q^2)}$. It diverges at the location of the point charge ($\bar{r}=\bar{r}_0=(\kappa q^2)^{1/4}$). We note that the non-singular feature of the strength of the electric field 
remains even in the solution of EiBI gravity coupled to Maxwell's electrodynamics \cite{banados,wei} and this is due to the Born-Infeld like structure in the gravity sector. However, this does not ensure the nonsingularity of the spacetime. In \cite{banados,wei}, the charge is not point-like, but distributed over a spherical surface.  $\rho_{obs}$ remains finite, though the scalar invariants diverge at this surface. But, in our solution, both the energy density and the curvature scalar are singular. A non-linear correlation between the gravity and the physical matter sector in the EiBI theory is thus revealed.     

\subsection{ Null geodesics}

Let us now turn to the null geodesics in this spacetime. On the equatorial 
plane ($\theta=\pi/2$), for null geodesics, we have, from Eq.~(\ref{eq:phyline_alpha1}),
\begin{equation}
\frac{\dot{r}^2}{2}\left(1-\frac{\kappa q^2}{\bar{r}^4}\right)+\frac{L^2}{2r^2(\bar{r})}\left(1+\frac{\kappa q^2}{\bar{r}^4} \right)\left(1-\frac{2M}{\bar{r}}+\frac{q^2}{\bar{r}^2} \right)=\frac{E^2}{2}
\label{eq:null_radial_eqn}
\end{equation} 
where, $E=\left(1+\frac{\kappa q^2}{\bar{r}^4} \right)\left(1-\frac{2M}{\bar{r}}+\frac{q^2}{\bar{r}^2} \right)\dot{t}$, $L=r^2(\bar{r})\dot{\phi} $ are two conserved quantities and $\dot{t},\, \dot{\phi},\, \dot{r}$ are defined as $\dot{t}=\frac{dt}{d\lambda},\,\dot{\phi}=\frac{d\phi}{d\lambda},\, \dot{r}=\frac{dr}{d\bar{r}}\frac{d\bar{r}}{d\lambda}$, $\lambda$ is the affine parameter. So, from Eq.~(\ref{eq:null_radial_eqn}) we identify the effective potential for the null geodesics as
\begin{equation}
V_{eff}(r)=\frac{L^2}{2r^2}\left(1+\frac{\kappa q^2}{\bar{r}^4} \right)\left(1-\frac{2M}{\bar{r}}+\frac{q^2}{\bar{r}^2} \right)
\label{eq:Veff_alpha1}
\end{equation}
where, $\bar{r}^2=(r^2+\sqrt{r^4+4\kappa q^2})/2$ .
The aforementioned effective potential is useful to study the propagation of
gravitational signals in the current scenario. We discuss photon propagation 
in the following subsection.

\subsection{ The propagation of photons}
Photons (lightlike particles) in Born-Infeld electromagnetism (BI photon), 
no longer propagate along the null geodesics of the background spacetime 
geometry. This is generally true for nonlinear electrodynamics \cite{plebansky}. Instead, the BI photon trajectories are determined by the null geodesics of 
the so-called effective geometry, given by \cite{plebansky,novello},
\begin{equation}
g^{\mu\nu}_{eff}=\left(1+\frac{1}{b^2}F\right)g^{\mu\nu}+\frac{1}{b^2}F^{\mu}_{\; \sigma}F^{\sigma \nu}
\label{eq:effective_metric}
\end{equation} 
where, $F=\frac{1}{2}F_{\alpha\beta}F^{\alpha\beta}$ and $g_{\mu\nu}$ defines the metric of the background spacetime. Such an effective geometry is induced due to nonlinear self-interaction of the photon. Here, in Eq.~(\ref{eq:effective_metric}), we assume the same $F_{\mu\nu}$ which generates background spacetime geometry. Then, the effective line element becomes
\begin{equation}
ds^2_{eff}=-\left(1+\frac{\kappa q^2}{\bar{r}^4}\right)\left(1-\frac{2M}{\bar{r}}+\frac{q^2}{\bar{r}^2}\right)dt^2+\frac{\left(1+\frac{\kappa q^2}{\bar{r}^4}\right)}{\left(1-\frac{2M}{\bar{r}}+\frac{q^2}{\bar{r}^2}\right)}d\bar{r}^2+\frac{\left(1+\frac{\kappa q^2}{\bar{r}^4}\right)^2}{\left(1-\frac{\kappa q^2}{\bar{r}^4}\right)}\bar{r}^2\left(d\theta^2 +\sin^2 \theta d \phi^2\right)
\label{eq:eff_line}
\end{equation}
Since our interest is to study the null geodesics of the effective geometry, we take out the conformal factor $\left(1+\frac{\kappa q^2}{\bar{r}^4}\right)$ and rewrite the effective line element as,
\begin{equation}
ds^2_{eff}=-\left(1-\frac{2M}{\bar{r}}+\frac{q^2}{\bar{r}^2}\right)dt^2+\frac{d\bar{r}^2}{\left(1-\frac{2M}{\bar{r}}+\frac{q^2}{\bar{r}^2}\right)}+\frac{\left(1+\frac{\kappa q^2}{\bar{r}^4}\right)}{\left(1-\frac{\kappa q^2}{\bar{r}^4}\right)}\bar{r}^2\left(d\theta^2 +\sin^2 \theta d \phi^2\right)
\label{eq:eff_line2}
\end{equation}
The effective geometry becomes exactly the 
same as the Reissner-Nordstr\"om solution when $\kappa=0$, which is expected. 
The analysis of the squared area radius in the effective geometry, 
$\left[r^2=\bar{r}^2\left(\frac{\bar{r}^4+\kappa q^2}{\bar{r}^4-\kappa q^2}\right)\right]$, 
shows that the area radius, or, the physical radial distance $r$ becomes infinity both at $\bar{r}=\bar{r}_0=(\kappa q^2)^{1/4}$ and $\bar{r}\rightarrow \infty$, but has a minimum 
$r_{th}^2=\frac{\sqrt{\kappa}q}{(\sqrt{5}-2)^{1/2}}\left(\frac{\sqrt{5}-1}{3-\sqrt{5}}\right)$ for $\bar{r}_{th}=\left(\frac{\kappa q^2}{\sqrt{5}-2}\right)^{1/4}$ (similar to the throat radius square of a wormhole). This implies that for 
each physical radial distance $r\in [r_{th},\infty]$, there exists a $\bar{r}\in [\bar{r}_{th},\infty]$ and 
a $\bar{r}\in [\bar{r}_0,\bar{r}_{th}]$. But these two sets of ranges of  
$\bar{r}$ describe two different geometries. We are interested in studying 
the propagation of photons coming from and going towards 
asymptotically flat regions of the given spacetime. Hence, we focus on 
$\bar{r}_{th}\leq \bar{r} < \infty$, which describes an asymptotically flat geometry.  

Now, for the null geodesics, we have
\begin{equation}
\frac{1}{2}\dot{\bar{r}}^2+ \frac{L^2}{2\bar{r}^2}\left(1-\frac{2M}{\bar{r}}+\frac{q^2}{\bar{r}^2}\right)\left(\frac{\bar{r}^4-\kappa q^2}{\bar{r}^4+\kappa q^2}\right)=\frac{E^2}{2}
\label{eq:null_photon}
\end{equation} 
where, $E=\left(1-\frac{2M}{\bar{r}}+\frac{q^2}{\bar{r}^2}\right)\dot{t}$ and $L=r^2(\bar{r})\dot{\phi} $ are two conserved quantities. One can also rewrite the Eq.~(\ref{eq:null_photon}) in terms of physical radial distance $r$ using the relation $r^2(\bar{r})=\bar{r}^2\left(\frac{\bar{r}^4+\kappa q^2}{\bar{r}^4-\kappa q^2}\right)$.  So, the effective potential for the BI photons become
\begin{equation}
V_{eff}(\bar{r})=\frac{L^2}{2\bar{r}^2}\left(1-\frac{2M}{\bar{r}}+\frac{q^2}{\bar{r}^2}\right)\left(\frac{\bar{r}^4-\kappa q^2}{\bar{r}^4+\kappa q^2}\right)
\label{eq:Veff_photon}
\end{equation} 

\begin{figure}[!htbp]
\centering
\subfigure[]{\includegraphics[width=2.0in,angle=360]{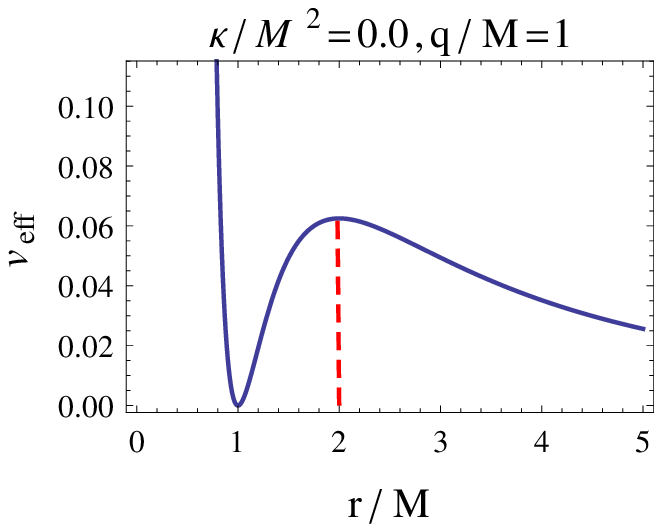}\label{subfig:K=0}}
\subfigure[]{\includegraphics[width=2.0in,angle=360]{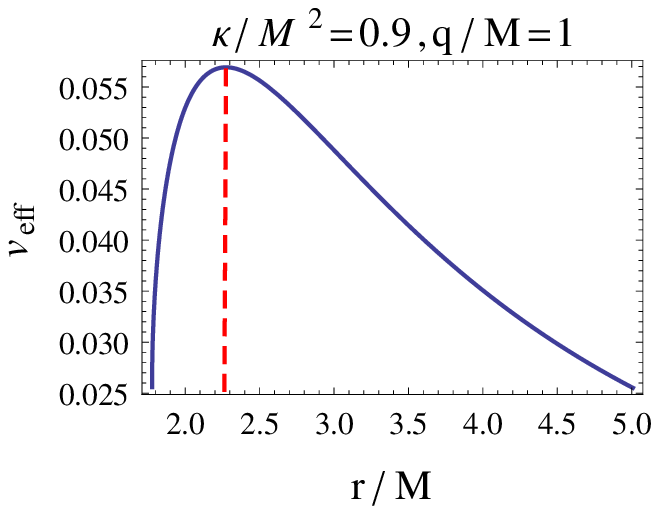}\label{subfig:K=0.9}}
\subfigure[]{\includegraphics[width=2.0in,angle=360]{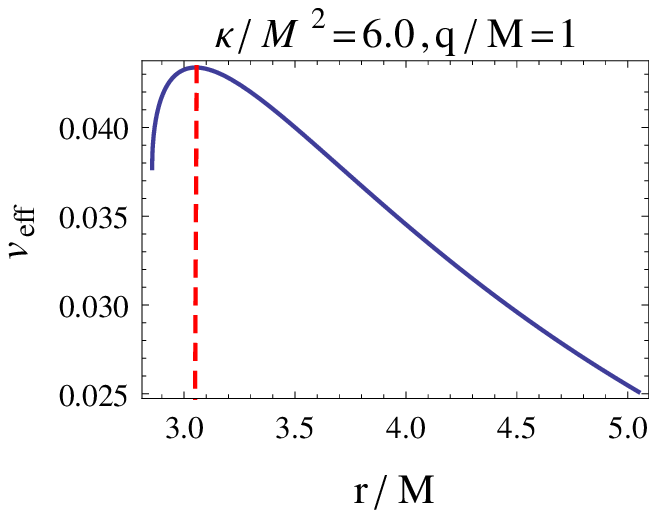}\label{subfig:K=6.0}}
\subfigure[]{\includegraphics[width=2.0in,angle=360]{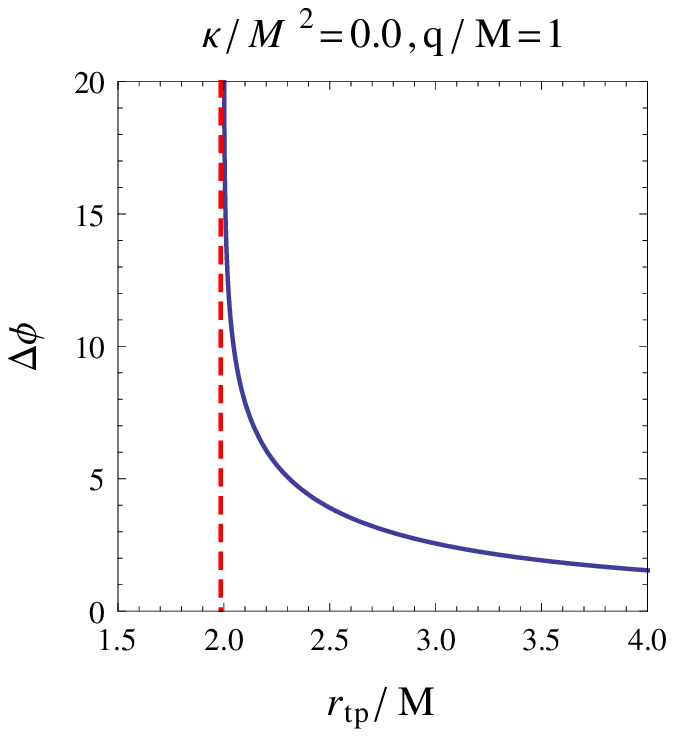}\label{subfig:angle1}}
\subfigure[]{\includegraphics[width=2.0in,angle=360]{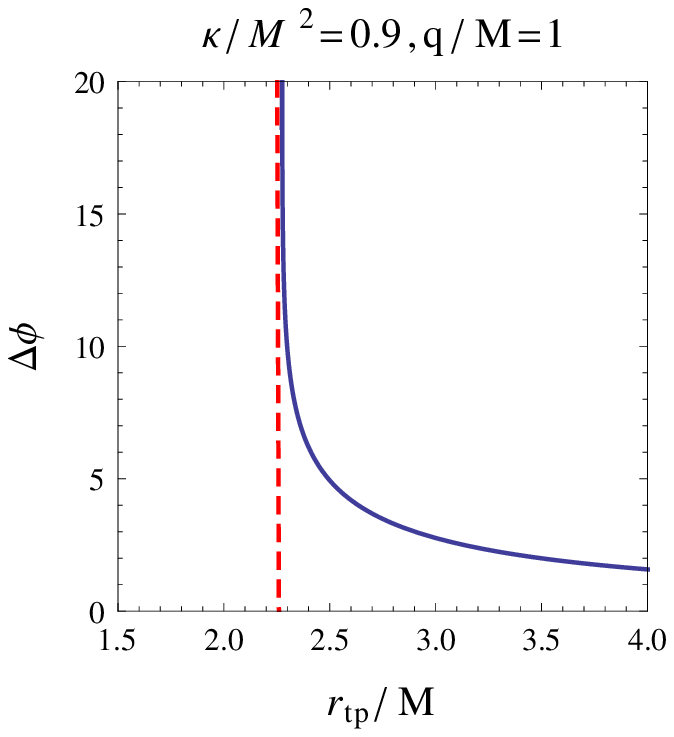}\label{subfig:angle2}}
\subfigure[]{\includegraphics[width=2.0in,angle=360]{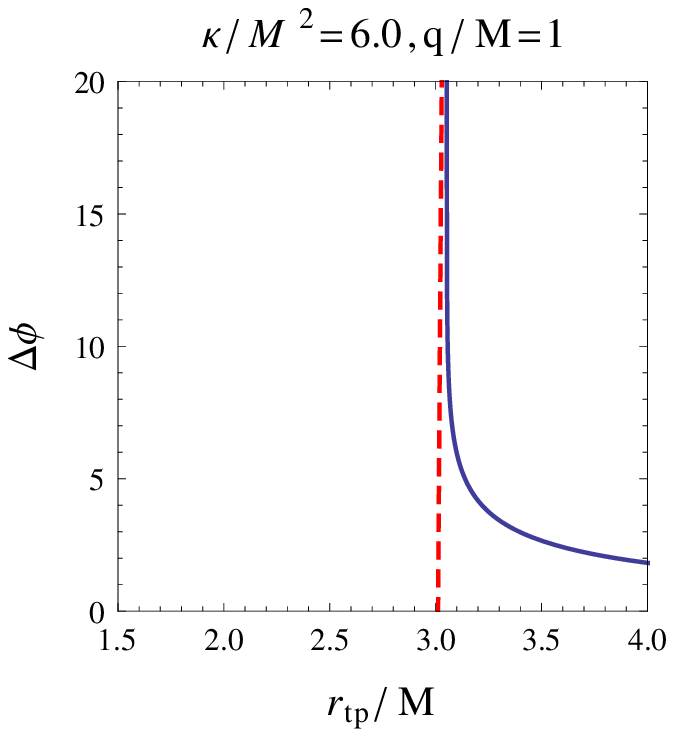}\label{subfig:angle3}}
\caption{(Top  panel)The effective potential [Eq.~(\ref{eq:Veff_photon})] is plotted as a function of the dimensionless quantity $r/M$, where $r(\bar{r})$ is the radius of the 2-sphere labelled by the radial coordinate $\bar{r}$   and $M$ is the mass. In the parametric plot,  the ($\bar{r}$) values are in the range $\bar{r}_{th}\leq \bar{r} < \infty$. In the plot, $v_{eff}={2M^2V_{eff}}/L^2$. The parameter values are specified on the top of the figures. The vertical red dashed line in each plot shows the correspondence between the photon sphere radius and the local maxima of the effective potential.  (Bottom panel)The exact deflection angle ($\Delta \phi$) is plotted as a function of $r_{tp}/M$, where $r_{tp}$ is the physical radial distance at the turning point. In each figure, $\Delta \phi$ rises sharply along the vertical red dashed line. The corresponding $r_{tp}/M$ relates to the photon sphere radius.}
\label{fig:deflection}
\end{figure}

We plot the effective potential for different values of 
$\kappa/M^2$ (a dimensionless parameter) and extract the essential features. 
A few plots are shown in the top panel of the Fig.~\ref{fig:deflection}. In these plots, the effective potential is rescaled as $v_{eff}={2M^2V_{eff}}/L^2$ and is plotted as a function of the dimensionless quantity $r/M$. In these parametric plots, the parameter ($\bar{r}$) value ranges from $\bar{r}_{th}$ to infinity. The maxima of the effective potential correspond to the unstable circular 
orbits for photons. The corresponding spherical surface is called the `photon sphere' \cite{wald,virbhadra3,virbhadra4}. In the Reissner-Nordstr\"om solution, the radius of the photon sphere depends on $q$ and $M$. In our solution, we note that, it depends additionally on $\kappa$. As an example, we consider the case for $q/M=1$. In the 
Reissner-Nordstr\"om solution, the radius of this photon sphere is 
$r_{ps}=2M=2q$ (see Fig.~\ref{subfig:K=0}). In our solution, we have a different picture. Here, from the figures of the top panel of Fig.~\ref{fig:deflection}, we see that as $\kappa$ is increased the radius of the photon sphere increases. This feature becomes more clear when we look at the plot of deflection angles ($\Delta \phi$) in the bottom panel of Fig.~\ref{fig:deflection}. Near photon sphere, the deflection angle becomes infinitely large, indicating multiple full rounds of light rays around the lensing object.  Consequently, a sequence of a large number of highly demagnified relativistic images may arise \cite{virbhadra1,virbhadra2}.

Analysing the plots of the effective potential for greater values of $\kappa/M^2$, one can easily verify that the radius of the photon sphere ($r_{ps}$) increases but becomes closer and closer to the throat radius ($r_{th}$) as $\kappa/M^2$-value increases.
\\

{\em Strong deflection}:
In this effective geometry, the exact expression for the deflection angle \cite{wald} 
becomes,
\begin{eqnarray}
\Delta \phi &=& 2\vert \phi(\infty)-\phi(r_{tp})\vert -\pi \nonumber \\
&=& 2 \int^{\infty}_{\bar{r}_{tp}}\frac{(\bar{r}^4-\kappa q^2)}{\bar{r}^2(\bar{r}^4+\kappa q^2)}\left[\frac{1}{\bar{r}^2_{tp}}\left(\frac{\bar{r}^4_{tp}-\kappa q^2}{\bar{r}^4_{tp}+\kappa q^2} \right)\left(1-\frac{2M}{\bar{r}_{tp}}+\frac{q^2}{\bar{r}_{tp}^2} \right)- \frac{1}{\bar{r}^2}\left(\frac{\bar{r}^4-\kappa q^2}{\bar{r}^4+\kappa q^2} \right)\left(1-\frac{2M}{\bar{r}}+\frac{q^2}{\bar{r}^2} \right)\right]^{-1/2}d\bar{r} \nonumber\\
&&-\pi
\label{eq:deflection_angle}
\end{eqnarray} 
where, $r^2(\bar{r}_{tp})=\bar{r}_{tp}^2\frac{\left(\bar{r}^4_{tp}+\kappa q^2 \right)}{\left(\bar{r}^4_{tp}-\kappa q^2 \right)}$. Here, $r_{tp}$ indicates the turning point, $i.e.$ $\frac{dr}{d\phi}\vert_{r_{tp}}=\frac{\dot{r}}{\dot{\phi}}\vert_{r_{tp}}=0$. The deflection angle is plotted as a function of the 
dimensionless parameter $r_{tp}/M$ and the figures are shown in the bottom panel 
of Fig.~\ref{fig:deflection}. Fig.~\ref{subfig:angle1} (for $\kappa/M^2=0$, $q/M=1$) shows that, 
in the extremal Reissner-Nordstr\"om case, the deflection angle increases 
rapidly ($>> 2\pi$) near the photon sphere ($r_{tp}=r_{ps}=2M$). From the next 
two figures (Fig.~\ref{subfig:angle2} and Fig.~\ref{subfig:angle3}), we see that the nature of plots do not change. However, as 
the $\kappa/M^2$ value increases, it is clearly seen that radius of the photon sphere 
increases from  the value $2M$.\\

{\em Weak deflection}: Let us now look for the leading order contribution 
of $\kappa$ in the weak-deflection angle formula. We compare this result 
with the Reissner-Nordstr\"om case. We obtain the weak deflection angle 
perturbatively \cite{bodenner}. For a spherically symmetric static line element, $ds^2=-A(\bar{r})dt^2+B(\bar{r})d\bar{r}^2+\bar{r}^2C(\bar{r})d\Omega^2$, the radial geodesic equation is
\begin{equation}
2B\ddot{\bar{r}}+B'\dot{\bar{r}}^2+A'\dot{t}^2-\left(\bar{r}^2C\right)'\dot{\phi}^2=0
\label{eq:ansatz_sphsymm}
\end{equation} 
With the conserved quantities, $E=A\dot{t}$, $L=C\bar{r}^2\dot{\phi} $, and the new variable $u=\frac{1}{\bar{r}}$, for the null geodesics, the Eq.~(\ref{eq:ansatz_sphsymm}) becomes \cite{bodenner}
\begin{equation}
\frac{d^2u}{d\phi^2}+\frac{C}{B}u=-\frac{1}{2}u^2\frac{d}{du}\left( \frac{C}{B}\right)+\frac{1}{2b^2_0}\frac{d}{du}\left(\frac{C^2}{AB}\right)
\label{eq:u-phi}
\end{equation}  
where, $b_0=L/E$ is the impact parameter. In the Reissner-Nordstr\"om case, 
$\left(A=B^{-1}=1-2Mu+q^2u^2,\, C=1 \right)$, the Eq.~(\ref{eq:u-phi}) becomes,
\begin{equation}
\frac{d^2u}{d\phi^2}+u=3Mu^2-2q^2u^3
\end{equation}
In our case, $A=B^{-1}=\left(1-2Mu+q^2u^2 \right)$, and $C=(1+\kappa q^2 u^4)/(1-\kappa q^2 u^4)$. We assume small $\kappa$ ($\kappa q^2 u^4<< 1$) and write the Eq.~(\ref{eq:u-phi}) keeping only the first order terms in $\kappa$,
\begin{equation}
\frac{d^2u}{d\phi^2}+u=3Mu^2-\left(2 -\frac{8\kappa}{b^2_0}\right)q^2u^3-6\kappa q^2 u^5 +14\kappa q^2 M u^6 -8\kappa q^4 u^7
\label{eq:u-phi_alpha1}
\end{equation} 
We set $\epsilon=Mu_N$, where $u_N=1/\bar{r}_N$ and $b_0=r_N=\bar{r}_N\left(\frac{\bar{r}_N^4+\kappa q^2}{\bar{r}_N^4-\kappa q^2}\right)$  ($r_N$ is the 
Newtonian distance of closest approach). Further, we 
define a dimensionless variable $\xi=u/u_N$. Thereafter, 
we write the Eq.~(\ref{eq:u-phi_alpha1}) to second order in $\epsilon$,
\begin{equation}
\frac{d^2\xi}{d\phi^2}+\xi\approx 3\epsilon \xi^2 - \epsilon^2 \left( 2-\frac{8\kappa}{b_0^2} \right)\frac{q^2}{M^2}\xi^3
\label{eq:xi-phi_2ndorder}
\end{equation} 
Expressing $\xi$ as $\xi= \xi_0+\epsilon\, \xi_1 + \epsilon^2\, \xi_2 +......$, and substituting them  in the Eq.~(\ref{eq:xi-phi_2ndorder})
we find the following set of equations,
by collecting terms of different order:
\begin{eqnarray}
\frac{d^2\xi_0}{d\phi^2}&+&\xi_0=0 \label{eq:1storder}\\
\frac{d^2\xi_1}{d\phi^2}&+&\xi_1-3\xi^2_0=0 \label{eq:2ndorder}\\
\frac{d^2\xi_2}{d\phi^2}&+&\xi_2-6\xi_0\xi_1+2\left(1-\frac{4\kappa}{b_0^2}\right)\frac{q^2}{M^2}\xi^3_0=0
\label{eq:3rdorder} 
\end{eqnarray}
Solving the equations~(\ref{eq:1storder}), ($\ref{eq:2ndorder}$), and ($\ref{eq:3rdorder}$) we get the approximate inverse radial distance,
\begin{eqnarray}
u&\simeq&  u_N\cos \phi + Mu_N^2 \left(\frac{3}{2} -\frac{1}{2}\cos( 2\phi)\right)+ \nonumber \\
 & &\frac{3M^2u_N^3}{16}\left( \left(20 - 4(1-\frac{4\kappa}{b_0^2})\frac{q^2}{M^2}\right)\phi \sin \phi +\left(1+\frac{q^2}{3M^2}(1-\frac{4\kappa}{b_0^2}) \right)\cos (3\phi)\right) 
 \label{eq:inverse_rdistance}
\end{eqnarray}
Assuming,  $u\rightarrow 0$ at $\phi= \pi/2 +\delta$, we solve for $\delta$ using the Eq.~(\ref{eq:inverse_rdistance}). For a small deflection angle, we find

\begin{equation}
\delta \approx 2 M u_N + \frac{3\pi M^2 u_N^2}{8}\left[5 - (1-\frac{4\kappa}{b_0^2})\frac{q^2}{M^2} \right]
+\mathcal{O}(M^3u_N^3)
\end{equation}
The deflection angle is $\Delta \phi = 2 \delta$. To get the first order in $\kappa$ contribution in the deflection angle, we use $b_0\approx 1/u_N$. At the distance of closest 
approach, $\bar{r}_{tp}=1/u(0)$. Converting from $u_N$ to $u(0)$, substituting 
$u(0)=1/\bar{r}_{tp}$, with $\bar{r}_{tp}\approx r_{tp}\left(1-{\kappa q^2}/r^4_{tp}+\mathcal{O}(\kappa^2)\right)$, we finally obtain the approximate weak deflection angle formula as,
\begin{equation}
\Delta \phi_{weak} \simeq \frac{4M}{r_{tp}}+\frac{M^2}{r^2_{tp}}\left(\frac{15}{4}\pi -4 \right)-\frac{3\pi}{4}\left(1-\frac{4\kappa}{r^2_{tp}} \right)\frac{q^2}{r_{tp}^2}
\label{eq:weak_deflection_angle}
\end{equation} 
In this formula, the term involving $\kappa$ is the new contribution 
to weak deflection angle. If $\kappa=0$, the expression becomes the same as in the 
Reissner-Nordstr\"om case \cite{briet}. By extending the calculation upto the fourth order perturbation ($i.e.$ $\mathcal{O}(\epsilon^4)$), one can verify that the sign of the leading order in $\kappa$ term that appears in the fourth order correction ($1/r_{tp}^4$-term) in the deflection angle, is still positive. This term involving $\kappa$ increases the net deflection angle. As the value of $\kappa$ is increased, the 
deflection angle also increases further. Earlier, we have seen that, with the 
increase of $\kappa$, the radius of photon-sphere increases. Thus, the study of 
both strong and weak deflection seem to suggest an overall attractive effect of $\kappa$.


\section{Effective potential and photon propagation for $\alpha \neq 1$ solutions}

In this section we investigate the effective potential, photon sphere, and light propagation for $\alpha \neq 1$ solutions through numerical graph and plots. We also compare these with that for the Maxwell electric field by taking different values of $b^2$ (or, $\alpha$ with fixed $\kappa$).
      
 The effective metric for the BI photon propagating in the background spacetime which is the EiBI solution with Born-Infeld elecric field ($\alpha \neq 1$), is derived from Eq.~(\ref{eq:effective_metric}) using Eqs.~[\ref{eq:physical_metric_ansatz}, \ref{eq:dphi_dr}] as,
\begin{equation}
ds^2_{eff}=-U(\bar{r})e^{2\psi(\bar{r})}dt^2+U(\bar{r})e^{-2\psi(\bar{r})}d\bar{r}^2+\left(\frac{\alpha V^2(\bar{r})\bar{r}^4+4\kappa q^2}{\alpha V(\bar{r})\bar{r}^2}\right)(d\theta^2 +\sin^2 \theta d\phi^2)
\label{eq:effective_metric_genalpha}
\end{equation} 
where, $U(\bar{r})$, $V(\bar{r})$, and $e^{2\psi(\bar{r})}$ have the expressions given by the Eqs.~[\ref{eq:U_alpha}, \ref{eq:V_alpha}, and \ref{eq:e2psi_classii} or, \ref{eq:e2psi_classiv}]. Using Eq.~(\ref{eq:effective_metric_genalpha}), and following the same analysis as described in  section IV, the effective potential for BI photon (for $\alpha \neq 1$) is derived as,
\begin{eqnarray}
V_{eff}(\bar{r})&=&\frac{U(\bar{r})e^{2\psi(\bar{r})}L^2}{2r^2(\bar{r})} \label{eq:Veff_alpha_gen}\\
\mbox{where,}\,r^2(\bar{r})&=&\left(\frac{\alpha V^2(\bar{r})\bar{r}^4+4\kappa q^2}{\alpha V(\bar{r})\bar{r}^2}\right) \label{eq:rsquare_BIph_alphagen}.
\end{eqnarray}
Here, the effective potential as a function of the physical radial distance, $i.e.$ $V_{eff}(r)$, is to be determined in a parametric form as given by Eqs.~[(\ref{eq:Veff_alpha_gen}), (\ref{eq:rsquare_BIph_alphagen})].

\begin{figure}[!htbp]
\centering
\subfigure[$\kappa=1.0, L=1.0, q=0.7, M=1.0$]{\includegraphics[width=3.0in,angle=360]{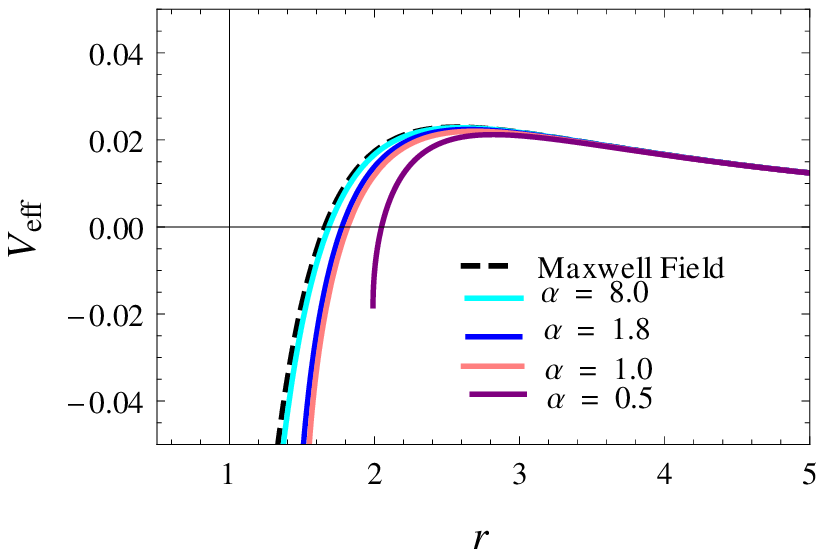}\label{subfig:Veff1}}
\subfigure[$\kappa=-1.0, L=1.0, q=0.9, M=1.0$]{\includegraphics[width=3.0in,angle=360]{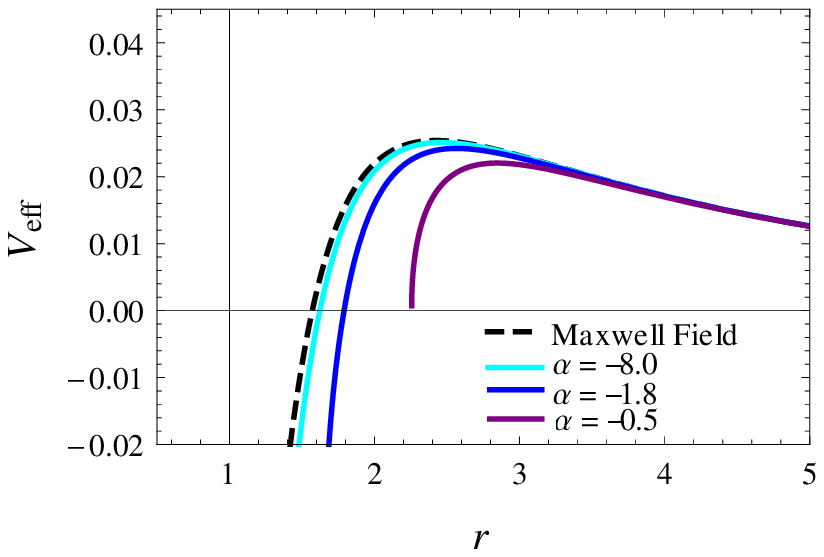}\label{subfig:Veff2}}
\caption{Effective potential ($V_{eff}$) for BI photons are plotted and compared with that of the Maxwellian photon (black dashed line) propagating in the background of EiBI spacetimes. In both plots, $\kappa$ value is same (but of opposite sign). However, $\alpha=4\kappa b^2$ takes different vaues in the plots. }
\label{fig:compare_veff}
\end{figure}

On the other hand, from Eq.~(\ref{eq:metric_kpos_maxwell}), we get the effective potential for the Maxwell photon as
\begin{equation}
V_{eff}^{\text{Max}}(r)=\frac{r^2h_1(r)L^2}{2(r^4-\kappa q^2)}
\label{eq:Veff_maxwell}
\end{equation}  
where, $h_1(r)$ is given by Eq.~(\ref{eq:h1}). In Eq.~(\ref{eq:Veff_maxwell}), the effective potential is written directly as a function of the physical radial distance $r$. For $\kappa <0$, $h_1(r)$ is replaced by $h_2(r)$ (given by Eq.~(\ref{eq:h2})). In Fig.~\ref{fig:compare_veff}, variation of effective potential for BI photons as compared to Maxwellian photon are shown. For plotting, we use Eqs.~[\ref{eq:Veff_maxwell},\ref{eq:Veff_alpha_gen}, \ref{eq:Veff_alpha1}]. Similar to the $\alpha=1$ special case, we consider $\bar{r} \in [\bar{r}_{th}, \infty] $, where, $\bar{r}_{th}^2= \sqrt{\kappa q^2/\alpha}[2+\alpha (\sqrt{2}-1)]$,  for a range of $\alpha$. From the plots in Fig.~\ref{fig:compare_veff}, it is evident that the radius of the photon sphere ($r_{ps}$) for BI photons is smaller than that of Maxwell photons (irrespective of positive or, negative $\kappa$). Furthermore, the photon sphere decreases as the values of $b^2$ decreases.  

\begin{figure}[!htbp]
\centering
\subfigure[$\kappa=1.0, L=1.0, q=0.7, M=1.0$]{\includegraphics[width=3.0in,angle=360]{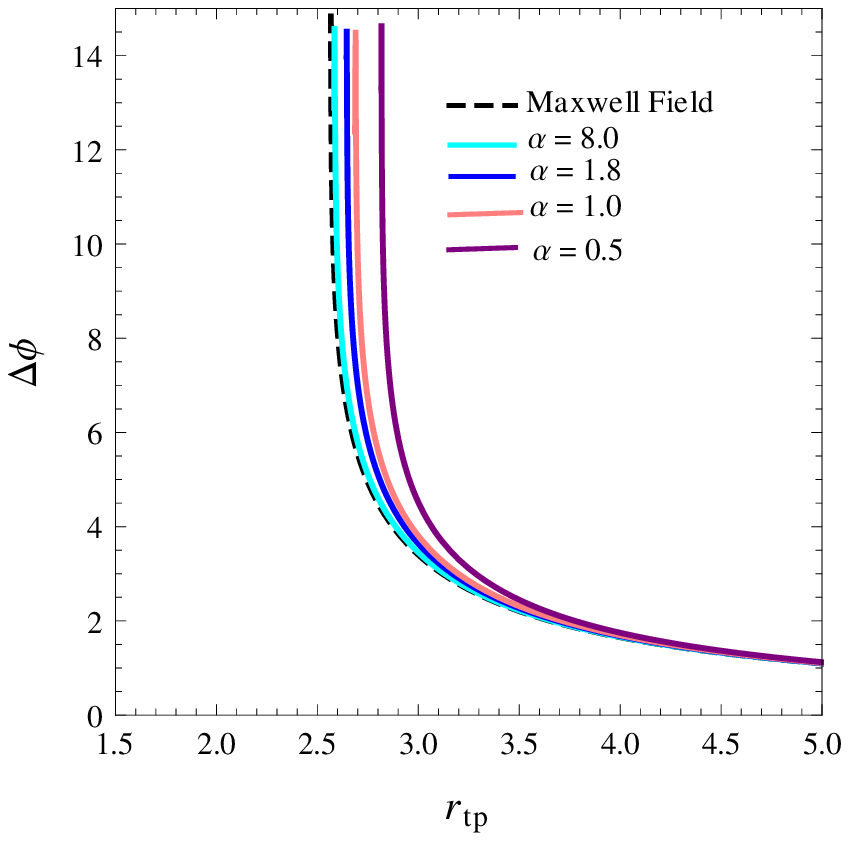}\label{subfig:deflection1}}
\subfigure[$\kappa=-1.0, L=1.0, q=0.9, M=1.0$]{\includegraphics[width=3.0in,angle=360]{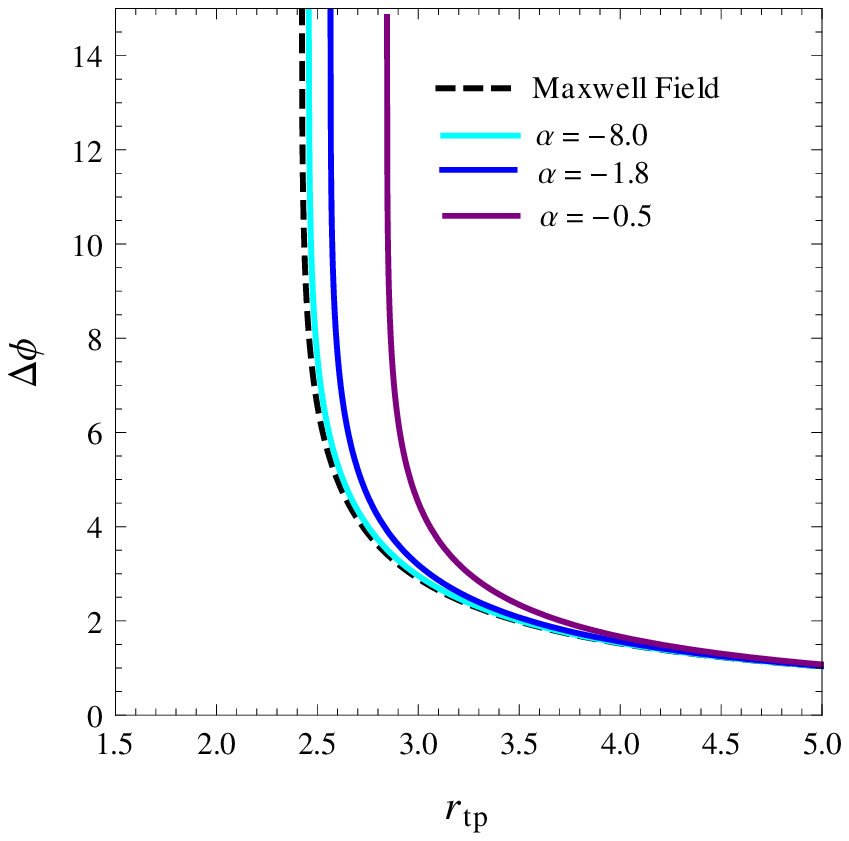}\label{subfig:deflection2}}
\caption{Deflection angle ($\Delta \phi$) for BI photons are plotted as a function of the physical radial distance at the turning point ($r_{tp}$) and compared with that of the Maxwellian photon (black dashed line) propagating in the background of EiBI spacetimes. The deflection angle diverges  rapidly when the turning-point-radius ($r_{tp}$) becomes closer to the radius of the photon sphere ($r_{ps}$). In both plots, the parameter values are the same as in Fig.~\ref{fig:compare_veff}.   }
\label{fig:compare_deflection_angle}
\end{figure}

Let us now write the expression for the deflection of light (BI photon) from the null geodesic equation for the effective metric given by Eq.~(\ref{eq:effective_metric_genalpha}). We get the expression for the deflection angle as,
\begin{equation}
\Delta \phi=2\int^{\infty}_{\bar{r}_{tp}}\frac{U(\bar{r})}{r^2(\bar{r})}\left[\frac{U(\bar{r}_{tp})e^{2\psi(\bar{r}_{tp})}}{r^2(\bar{r}_{tp})}-\frac{U(\bar{r})e^{2\psi(\bar{r})}}{r^2(\bar{r})}\right]^{-1/2}d\bar{r} -\pi
\label{eq:deflection_angle_genalpha}
\end{equation}  
where, $r^2(\bar{r})$ is the physical radial distance square given by Eq.~(\ref{eq:rsquare_BIph_alphagen}) and $\bar{r}_{tp}$ is the coordinate value corresponding to the physical radial distance at the turning point ($r_{tp}$). Again, for the Maxwell photon, we get the deflection angle,
\begin{equation}
\Delta \phi^{\text{Max}}=2\int^{\infty}_{r_{tp}}\frac{1}{\sqrt{r^4+\kappa q^2}}\left[\frac{r^2_{tp}h_1(r_{tp})}{r^4_{tp}-\kappa q^2}-\frac{r^2h_1(r)}{r^4-\kappa q^2}\right]^{-1/2}dr -\pi
\label{eq:deflection_angle_maxwell}
\end{equation} 
where (\ref{eq:deflection_angle_maxwell}) is expressed directly in terms of physical radial distance ($r$). In Fig.~\ref{fig:compare_deflection_angle}, the variation of the deflection angle of light as function of  $r_{tp}$ is shown for different $\alpha$ values and it is compared with that for the Maxwell field. To visualize the correspondence between the effective potential for photons with the deflection angle of light, same parameter values are used in the plots. For plotting, Eqs.~[\ref{eq:deflection_angle_genalpha}, \ref{eq:deflection_angle_maxwell}, \ref{eq:deflection_angle}] are used.

From the plots of both the effective potential (Fig.~\ref{fig:compare_veff}) and strong deflection angle (Fig.~\ref{fig:compare_deflection_angle}), we note that with the increase of the value of $b^2$, the radius of the photon sphere ($r_{ps}$) decreases.  

\section{Conclusions}
In this article, we have derived new classes of spherically symmetric static 
solutions in EiBI gravity coupled to Born-Infeld electrodynamics. 
The spacetimes demonstrate the effect of coupling a non-linear electrodynamics (Born-Infeld electrodynamics) to a modified theory of 
gravity (the EiBI gravity). All solutions are analytical. We summarize our 
results below, point wise:

$(i)$ We have used a non-standard choice for the physical and auxiliary metrics to find the spacetimes 
in EiBI gravity. 
The standard ansatze for the physical and auxiliary line elements 
as employed in \cite{banados,wei,sotani,eibiwormhole}, led to a 
complicated set of coupled differential equations. 
We note that our non-standard choice is far more useful insofar as solution
construction is concerned.
Similar unconventional choices may be used in other 
cases with the hope of simplifying the set of equations we need to solve.

$(ii)$ We obtain two classes of spacetimes in section III. All the known 
results for different electrically charged, spherically symmetric, static scenarios 
can be obtained by taking different limits of 
$\alpha=4\kappa b^2$ and $\kappa$. In general, the spacetimes we obtain 
are singular at the location of the charge, where the charge can be localized 
about a point or distributed over a spherical shell (2-sphere). 
The singularity may or may not be covered by an event horizon, i.e. it may be a
black hole or a naked singularity. However, all solutions converge to the 
Reissner-Nordstr\"om solution at large distances.

$(iii)$ The effects of modifications in both gravity and matter sectors are demonstrated through the analysis of the metric functions for fixed $b^2$, varying $\kappa$ and vice-versa. In our analysis, we find some interesting features. For a fixed $\kappa$, the horizon radius increases with the decrease of $b^2$ for some parameter values ($\kappa, q, M$). If $b^2$ is fixed, the horizon radius decreases with increasing $\kappa$ ($\kappa> 0$).       

$(iv)$ Although the results are analytical, the expressions for the metric
functions are complicated. However, a special choice, $\alpha=1$ ($\kappa>0$), 
leads to a simple solution that looks similar to the Reissner-Nordstr\"om. 
We analyze this special solution in detail in section IV. We find that this 
spacetime also possesses the same generic features mentioned above in ($ii$). 
Subsequently, for the $\alpha =1$ line element, we investigate the null geodesics in the effective geometry for BI photons, as well as gravitational lensing. We find that as $\kappa$ is 
increased, 
the radius of the photon sphere increases. 
We also find the leading order contribution of $\kappa$ in the expression for
the weak deflection angle. The overall sign of this term is positive but
it depends on the value of $\kappa$. 

$(v)$In section V we investigate the effective potential, photon sphere, and strong deflection angle for $\alpha \neq 1$, through numerics and plots. We compare with results for a Maxwellian electric field. Analyzing  the plots, we find that for a fix value of $\kappa$ (irrespective of its sign), the radius of the photon sphere decreases as the value of $b^2$ is increased. 

\

\noindent In summary, using non-standard ansatze for the physical and
auxiliary metrics, we have been able to obtain analytical solutions in Born-Infeld
gravity coupled to Born-Infeld electrodynamics. One may explore strong gravitational lensing further using the approach in \cite{bozza}.  It is possible that there are other non-singular solutions as well
which are non-singular from both the gravitational and electrodynamic
perspectives. For this, it is essential to scan the $(\kappa, b^2, \frac{q}{M})$ parameter space more thoroughly. In \cite{rajibul}, the author has shown that in the case of EiBI solutions with  a Maxwellian electric field, there exists a region in the parameter space ($\kappa, q, M$) which allows the construction of a Lorentzian wormhole without violating the weak or null energy conditions. In our solution, a similar analysis is possible. One can also study the black hole thermodyanmics using the solutions presented here. We hope to work on these issues and communicate 
our results in the near future.

\section*{Acknowledgments}
The authors thank Jonas P. Pereira for pointing out a crucial fact 
on light propagation in nonlinear electrodynamics and also for his
comments.

\appendix*

\section{Integral results}
Here, we compute the integrals $\int V(\bar{r})\bar{r}^2d\bar{r}$, in detail, which have been used in the section \ref{sec:classification} to obtain the analytical expressions of two different classes of spacetimes [Eq.~(\ref{eq:e2psi_classii}) and Eq.~(\ref{eq:e2psi_classiv})]. These integrals are non-trivial and may not be computed correctly otherwise (e.g. using {\em Mathematica}).\\
  
{\em (a)$\infty > \alpha >1 $ case}:
\begin{eqnarray}
\int V\bar{r}^2d\bar{r}= -\frac{(2-\alpha)\bar{r}^3}{6(\alpha -1)}+\frac{\alpha}{2(\alpha -1)}\int \sqrt{\bar{r}^4-\frac{4\kappa q^2(\alpha -1)}{\alpha}}d\bar{r} 
\label{eq:integralcase2}
\end{eqnarray}
The last term on the R.H.S. of the Eq.~\ref{eq:integralcase2}, is a non-trivial integral which we evaluate below:
\begin{eqnarray}
\quad~ \int\sqrt{\bar{r}^4-\beta}d\bar{r}&=&\bar{r}\sqrt{\bar{r}^4-\beta}-\int \frac{2\bar{r}^4d\bar{r}}{\sqrt{\bar{r}^4-\beta}}, \quad~ (\mbox{ Using partial integration})\nonumber \\
&=&   \bar{r}\sqrt{\bar{r}^4-\beta}-2\int\sqrt{\bar{r}^4-\beta}d\bar{r} -2\beta\int \frac{d\bar{r}}{\sqrt{\bar{r}^4-\beta}} \nonumber\\
&=& \frac{1}{3}\bar{r}\sqrt{\bar{r}^4-\beta}-\frac{2\beta^{3/4}}{3}\int \frac{dy}{\sqrt{y^4-1}}, \quad~ (\bar{r}=\beta^{1/4}y) \nonumber \\
\mbox{Now,}\quad~ \int \frac{dy}{\sqrt{y^4-1}}&=& -\int \frac{du}{\sqrt{1-u^4}}, \quad~ (u=1/y) \nonumber\\
&=&-F\left(\arcsin (1/y)\middle \vert-1 \right)+C_{ext} \nonumber\\
\mbox{So,} \quad~ \int\sqrt{\bar{r}^4-\beta}d\bar{r} &=& \frac{1}{3}\bar{r}\sqrt{\bar{r}^4-\beta} + \frac{2\beta^{3/4}}{3}\left(F\left(\arcsin (\beta^{1/4}/\bar{r})\middle \vert-1 \right)-C_{ext}\right)
\label{eq:int1}
\end{eqnarray}
where, $F(\phi\vert m)=\int^{\phi}_0[1-m\sin^2\theta]^{-1/2}d\theta$ is the incomplete elliptic integral of the first kind. Here,  $\beta=\frac{4\kappa q^2 (\alpha -1)}{\alpha}$ ($\beta>0$) and $C_{ext}$ is an additional constant term appearing due to the fact that Elliptic integrals are defined with the zero lower limit and hence, these are not  the so-called indefinite integrals. The Eq.~(\ref{eq:integralcase2}) becomes,
\begin{eqnarray}
\int V\bar{r}^2d\bar{r}= \frac{\bar{r}^3}{3}&+&\frac{\alpha \bar{r}^3}{6(\alpha -1)}\left[\sqrt{1-\frac{4\kappa q^2(\alpha -1)}{\alpha \bar{r}^4}}-1\right]+\nonumber\\
&&\frac{\alpha^{1/4}(4\kappa q^2)^{3/4}}{3(\alpha -1)^{1/4}}\left[F\left(\arcsin\left(\frac{\left(4\kappa q^2(\alpha -1)\right)^{1/4}}{\alpha^{1/4}\bar{r}}\right)\middle \vert -1\right) -C_{ext}\right]
\label{eq:vint1}
\end{eqnarray}
Now, to fix up the extra constant $C_{ext}$, we use the fact that, in the limit of $\alpha\rightarrow \infty$ and $\kappa \rightarrow 0$, the full expression of $e^{2\psi}$ must take form of the Reissner-Nordstr\"om solution (i.e. $e^{2\psi}\rightarrow 1-\frac{2M}{\bar{r}}+\frac{q^2}{\bar{r}^2}$). Under this constraint, $C_{ext}=0$.\\

{\em (b)$-\infty < \alpha <1 $ case}:   
\begin{eqnarray}
\int V\bar{r}^2d\bar{r}= -\frac{(2-\alpha)\bar{r}^3}{6(\alpha -1)}+\frac{\alpha}{2(\alpha -1)}\int \sqrt{\bar{r}^4+\gamma}d\bar{r} 
\label{eq:integralcase4}
\end{eqnarray}
where, $\gamma=\frac{4\kappa q^2(1-\alpha)}{\alpha}$ ($\gamma>0$). We evaluate the last integral on R.H.S. of the Eq.~(\ref{eq:integralcase4}) below:
\begin{eqnarray}
\int\sqrt{\bar{r}^4+\gamma}d\bar{r}&=& \frac{1}{3}r\sqrt{\bar{r}^4+\gamma}+\frac{2\gamma^{3/4}}{3}\int \frac{dy}{\sqrt{y^4+1}}, \quad~ (\bar{r}=\gamma^{1/4}y)
\nonumber\\
\mbox{Now,}\quad~ \int \frac{dy}{\sqrt{y^4+1}}&=& -\sqrt{i}F\left(i\,arcsinh(\sqrt{i}y)\middle \vert-1 \right)+C'_{ext} \nonumber\\
\mbox{So,}\quad~ \int\sqrt{\bar{r}^4+\gamma}d\bar{r}&=& \frac{1}{3}\bar{r}\sqrt{\bar{r}^4+\gamma} -\frac{2\gamma^{3/4}}{3}\left[\sqrt{i}F\left(i\, arcsinh(\sqrt{i}r\gamma^{-1/4})\middle \vert-1 \right)-C'_{ext}\right]
\label{eq:int2}
\end{eqnarray}
where, $i=\sqrt{-1}$.
So, the Eq.~(\ref{eq:integralcase4}) becomes
\begin{eqnarray}
\int V\bar{r}^2d\bar{r}&=&\frac{\bar{r}^3}{3}+\frac{\alpha \bar{r}^3}{6(\alpha -1)}\left[\sqrt{1+\frac{4\kappa q^2 (1-\alpha)}{\alpha \bar{r}^4}}-1 \right] \nonumber\\
&& +\frac{\alpha^{1/4}(4\kappa q^2)^{3/4}}{3(1-\alpha)^{1/4}}\left[\sqrt{i}F\left(i \mbox{arcsinh}\left(\frac{\sqrt{i}\alpha^{1/4}\bar{r}}{(4\kappa q^2(1-\alpha))^{1/4}}\right)\middle \vert-1\right)-C'_{ext}\right]
\label{eq:vint2}
\end{eqnarray}
Similar to the earlier case, we fix up the extra constant $C'_{ext}$, by taking the limit of $\alpha\rightarrow -\infty$ and $\kappa \rightarrow -0$. Then also, the full expression of $e^{2\psi}$ takes the form of the Reissner-Nordstr\"om solution. Under this constraint, $C'_{ext}=-\int^{\infty}_0\frac{dy}{\sqrt{y^4+1}}=-\frac{\Gamma^2 (1/4)}{4\sqrt{\pi}}$. Eq.~(\ref{eq:vint2}) can be rewritten further as,
\begin{equation}
\int V\bar{r}^2d\bar{r}=\frac{\bar{r}^3}{3}+\frac{\alpha \bar{r}^3}{6(\alpha -1)}\left[\sqrt{1+\frac{4\kappa q^2 (1-\alpha)}{\alpha \bar{r}^4}}-1 \right] +\left(\frac{4\kappa q^2}{3\bar{r}}\right) {}_{2}{F}_{1}\left(\frac{1}{4},\frac{1}{2};\frac{5}{4};-\frac{4\kappa q^2(1-\alpha)}{\alpha \bar{r}^4}\right)
\label{eq:vint2_hyper}
\end{equation}
where, we use
\begin{equation}
\sqrt{i}F\left(i \mbox{arcsinh}\left(\sqrt{i}y\right)\middle \vert-1\right)+\frac{\Gamma^2 (1/4)}{4\sqrt{\pi}}=\int^{\infty}_y\frac{dz}{\sqrt{z^4+1}}=\frac{ {}_{2}F_{1}\left(\frac{1}{4},\frac{1}{2};\frac{5}{4};-\frac{1}{y^4}\right)}{y}.
\nonumber
\end{equation}
Here, ${}_{2}F_1(a,b;c;z)$ is the hypergeometric function with usual meaning.
\bibliographystyle{apsrev4-1}
\bibliography{reference}

\begin{thebibliography}{10}%
\makeatletter
\providecommand \@ifxundefined [1]{%
 \ifx #1\undefined \expandafter \@firstoftwo
 \else \expandafter \@secondoftwo
\fi
}%
\providecommand \@ifnum [1]{%
 \ifnum #1\expandafter \@firstoftwo
 \else \expandafter \@secondoftwo
\fi
}%
\providecommand \enquote [1]{``#1''}%
\providecommand \bibnamefont  [1]{#1}%
\providecommand \bibfnamefont [1]{#1}%
\providecommand \citenamefont [1]{#1}%
\providecommand\href[0]{\@sanitize\@href}%
\providecommand\@href[1]{\endgroup\@@startlink{#1}\endgroup\@@href}%
\providecommand\@@href[1]{#1\@@endlink}%
\providecommand \@sanitize [0]{\begingroup\catcode`\&12\catcode`\#12\relax}%
\@ifxundefined \pdfoutput {\@firstoftwo}{%
 \@ifnum{\z@=\pdfoutput}{\@firstoftwo}{\@secondoftwo}%
}{%
 \providecommand\@@startlink[1]{\leavevmode\special{html:<a href="#1">}}%
 \providecommand\@@endlink[0]{\special{html:</a>}}%
}{%
 \providecommand\@@startlink[1]{%
  \leavevmode
  \pdfstartlink
   attr{/Border[0 0 1 ]/H/I/C[0 1 1]}%
   user{/Subtype/Link/A<</Type/Action/S/URI/URI(#1)>>}%
  \relax
 }%
 \providecommand\@@endlink[0]{\pdfendlink}%
}%
\providecommand \url  [0]{\begingroup\@sanitize \@url }%
\providecommand \@url [1]{\endgroup\@href {#1}{\urlprefix}}%
\providecommand \urlprefix [0]{URL }%
\providecommand \Eprint[0]{\href }%
\@ifxundefined \urlstyle {%
  \providecommand \doi [1]{doi:\discretionary{}{}{}#1}%
}{%
  \providecommand \doi [0]{doi:\discretionary{}{}{}\begingroup
  \urlstyle{rm}\Url }%
}%
\providecommand \doibase [0]{http://dx.doi.org/}%
\providecommand \Doi[1]{\href{\doibase#1}}%
\providecommand \bibAnnote [3]{%
  \BibitemShut{#1}%
  \begin{quotation}\noindent
    \textsc{Key:}\ #2\\\textsc{Annotation:}\ #3%
  \end{quotation}%
}%
\providecommand \bibAnnoteFile [2]{%
  \IfFileExists{#2}{\bibAnnote {#1} {#2} {\input{#2}}}{}%
}%
\providecommand \typeout [0]{\immediate \write \m@ne }%
\providecommand \selectlanguage [0]{\@gobble}%
\providecommand \bibinfo [0]{\@secondoftwo}%
\providecommand \bibfield [0]{\@secondoftwo}%
\providecommand \translation [1]{[#1]}%
\providecommand \BibitemOpen[0]{}%
\providecommand \bibitemStop [0]{}%
\providecommand \bibitemNoStop [0]{.\EOS\space}%
\providecommand \EOS [0]{\spacefactor3000\relax}%
\providecommand \BibitemShut [1]{\csname bibitem#1\endcsname}%
\bibitem{hawk}%
  \BibitemOpen
  \bibfield{author}{%
  \bibinfo {author} {\bibfnamefont{S.~W.}\ \bibnamefont{Hawking}}\ and\
  \bibinfo {author} {\bibfnamefont{G.~F.~R.}\ \bibnamefont{Ellis}},\ }%
  \emph{\bibinfo {title} {The Large Scale Structure of Space-Time}}\ (\bibinfo
  {publisher} {Cambridge University Press},\ \bibinfo {address} {Cambridge,
  England},\ \bibinfo {year} {1975})%
  \bibAnnoteFile{NoStop}{hawk}%
\bibitem{edd}%
  \BibitemOpen
  \bibfield{author}{%
  \bibinfo {author} {\bibfnamefont{A.}~\bibnamefont{Eddington}},\ }%
  \emph{\bibinfo {title} {The Mathematical Theory of Relativity}}\ (\bibinfo
  {publisher} {Cambridge University Press},\ \bibinfo {address} {Cambridge,
  England},\ \bibinfo {year} {1924})%
  \bibAnnoteFile{NoStop}{edd}%
\bibitem{born}%
  \BibitemOpen
  \bibfield{author}{%
  \bibinfo {author} {\bibfnamefont{M.}~\bibnamefont{Born}}\ and\ \bibinfo
  {author} {\bibfnamefont{L.}~\bibnamefont{Infeld}},\ }%
  \bibfield{journal}{%
  \bibinfo {journal} {Proc. R. Soc. A}\ }%
  \textbf{\bibinfo {volume} {144}},\ \bibinfo {pages} {425} (\bibinfo {year}
  {1934})%
  \bibAnnoteFile{NoStop}{born}%
\bibitem{desgib}%
  \BibitemOpen
  \bibfield{author}{%
  \bibinfo {author} {\bibfnamefont{S.}~\bibnamefont{Deser}}\ and\ \bibinfo
  {author} {\bibfnamefont{G.~W.}\ \bibnamefont{Gibbons}},\ }%
  \bibfield{journal}{%
  \bibinfo {journal} {Classical and Quantum Gravity}\ }%
  \textbf{\bibinfo {volume} {15}},\ \bibinfo {pages} {L35} (\bibinfo {year}
  {1998})%
  \bibAnnoteFile{NoStop}{desgib}%
\bibitem{vollick}%
  \BibitemOpen
  \bibfield{author}{%
  \bibinfo {author} {\bibfnamefont{D.~N.}\ \bibnamefont{Vollick}},\ }%
  \bibfield{journal}{%
  \Doi{10.1103/PhysRevD.69.064030}{\bibinfo {journal} {Phys. Rev. D}}\ }%
  \textbf{\bibinfo {volume} {69}},\ \bibinfo {pages} {064030} (\bibinfo {year}
  {2004})%
  \bibAnnoteFile{NoStop}{vollick}%
\bibitem{vollick2}%
  \BibitemOpen
  \bibfield{author}{%
  \bibinfo {author} {\bibfnamefont{D.~N.}\ \bibnamefont{Vollick}},\ }%
  \bibfield{journal}{%
  \Doi{10.1103/PhysRevD.72.084026}{\bibinfo {journal} {Phys. Rev. D}}\ }%
  \textbf{\bibinfo {volume} {72}},\ \bibinfo {pages} {084026} (\bibinfo {year}
  {2005})%
  \bibAnnoteFile{NoStop}{vollick2}%
\bibitem{vollick3}%
  \BibitemOpen
  \bibfield{author}{%
  \bibinfo {author} {\bibfnamefont{D.~N.}\ \bibnamefont{{Vollick}}},\ }%
  \Eprint{http://arxiv.org/abs/arXiv:gr-qc/0601136}{arXiv:gr-qc/0601136}%
  \bibAnnoteFile{NoStop}{vollick3}%
\bibitem{banados}%
  \BibitemOpen
  \bibfield{author}{%
  \bibinfo {author} {\bibfnamefont{M.}~\bibnamefont{Ba\~nados}}\ and\ \bibinfo
  {author} {\bibfnamefont{P.~G.}\ \bibnamefont{Ferreira}},\ }%
  \bibfield{journal}{%
  \Doi{10.1103/PhysRevLett.105.011101}{\bibinfo {journal} {Phys. Rev. Lett.}}\
  }%
  \textbf{\bibinfo {volume} {105}},\ \bibinfo {pages} {011101} (\bibinfo {year}
  {2010})%
  \bibAnnoteFile{NoStop}{banados}%
\bibitem{isham}%
  \BibitemOpen
  \bibfield{author}{%
  \bibinfo {author} {\bibfnamefont{C.~J.}\ \bibnamefont{Isham}}, \bibinfo
  {author} {\bibfnamefont{A.}~\bibnamefont{Salam}},\ and\ \bibinfo {author}
  {\bibfnamefont{J.}~\bibnamefont{Strathdee}},\ }%
  \bibfield{journal}{%
  \Doi{10.1103/PhysRevD.3.867}{\bibinfo {journal} {Phys. Rev. D}}\ }%
  \textbf{\bibinfo {volume} {3}},\ \bibinfo {pages} {867} (\bibinfo {year}
  {1971})%
  \bibAnnoteFile{NoStop}{isham}%
\bibitem{scargill}%
  \BibitemOpen
  \bibfield{author}{%
  \bibinfo {author} {\bibfnamefont{J.~H.~C.}\ \bibnamefont{Scargill}}, \bibinfo
  {author} {\bibfnamefont{M.}~\bibnamefont{Banados}},\ and\ \bibinfo {author}
  {\bibfnamefont{P.~G.}\ \bibnamefont{Ferreira}},\ }%
  \bibfield{journal}{%
  \Doi{10.1103/PhysRevD.86.103533}{\bibinfo {journal} {Phys. Rev. D}}\ }%
  \textbf{\bibinfo {volume} {86}},\ \bibinfo {pages} {103533} (\bibinfo {year}
  {2012})%
  \bibAnnoteFile{NoStop}{scargill}%
\bibitem{schmidt}%
  \BibitemOpen
  \bibfield{author}{%
  \bibinfo {author} {\bibfnamefont{A.}~\bibnamefont{Schmidt-May}}\ and\
  \bibinfo {author} {\bibfnamefont{M.}~\bibnamefont{von Strauss}}}%
   (\bibinfo {year} {2014}),\
  \Eprint{http://arxiv.org/abs/1412.3812}{arXiv:1412.3812 [hep-th]}%
  \bibAnnoteFile{NoStop}{schmidt}%
\bibitem{lavinia}%
  \BibitemOpen
  \bibfield{author}{%
  \bibinfo {author} {\bibfnamefont{J.~B.}\ \bibnamefont{Jiménez}}, \bibinfo
  {author} {\bibfnamefont{L.}~\bibnamefont{Heisenberg}},\ and\ \bibinfo
  {author} {\bibfnamefont{G.~J.}\ \bibnamefont{Olmo}},\ }%
  \bibfield{journal}{%
  \bibinfo {journal} {Journal of Cosmology and Astroparticle Physics}\ }%
  \textbf{\bibinfo {volume} {2014}},\ \bibinfo {pages} {004} (\bibinfo {year}
  {2014})%
  \bibAnnoteFile{NoStop}{lavinia}%
\bibitem{cardoso}%
  \BibitemOpen
  \bibfield{author}{%
  \bibinfo {author} {\bibfnamefont{P.}~\bibnamefont{Pani}}, \bibinfo {author}
  {\bibfnamefont{V.}~\bibnamefont{Cardoso}},\ and\ \bibinfo {author}
  {\bibfnamefont{T.}~\bibnamefont{Delsate}},\ }%
  \bibfield{journal}{%
  \Doi{10.1103/PhysRevLett.107.031101}{\bibinfo {journal} {Phys. Rev. Lett.}}\
  }%
  \textbf{\bibinfo {volume} {107}},\ \bibinfo {pages} {031101} (\bibinfo {year}
  {2011})%
  \bibAnnoteFile{NoStop}{cardoso}%
\bibitem{casanellas}%
  \BibitemOpen
  \bibfield{author}{%
  \bibinfo {author} {\bibfnamefont{J.}~\bibnamefont{Casanellas}}, \bibinfo
  {author} {\bibfnamefont{P.}~\bibnamefont{Pani}}, \bibinfo {author}
  {\bibfnamefont{I.}~\bibnamefont{Lopes}},\ and\ \bibinfo {author}
  {\bibfnamefont{V.}~\bibnamefont{Cardoso}},\ }%
  \bibfield{journal}{%
  \bibinfo {journal} {The Astrophysical Journal}\ }%
  \textbf{\bibinfo {volume} {745}},\ \bibinfo {pages} {15} (\bibinfo {year}
  {2012})%
  \bibAnnoteFile{NoStop}{casanellas}%
\bibitem{avelino}%
  \BibitemOpen
  \bibfield{author}{%
  \bibinfo {author} {\bibfnamefont{P.~P.}\ \bibnamefont{Avelino}},\ }%
  \bibfield{journal}{%
  \Doi{10.1103/PhysRevD.85.104053}{\bibinfo {journal} {Phys. Rev. D}}\ }%
  \textbf{\bibinfo {volume} {85}},\ \bibinfo {pages} {104053} (\bibinfo {year}
  {2012})%
  \bibAnnoteFile{NoStop}{avelino}%
\bibitem{sham}%
  \BibitemOpen
  \bibfield{author}{%
  \bibinfo {author} {\bibfnamefont{Y.-H.}\ \bibnamefont{Sham}}, \bibinfo
  {author} {\bibfnamefont{L.-M.}\ \bibnamefont{Lin}},\ and\ \bibinfo {author}
  {\bibfnamefont{P.~T.}\ \bibnamefont{Leung}},\ }%
  \bibfield{journal}{%
  \Doi{10.1103/PhysRevD.86.064015}{\bibinfo {journal} {Phys. Rev. D}}\ }%
  \textbf{\bibinfo {volume} {86}},\ \bibinfo {pages} {064015} (\bibinfo {year}
  {2012})%
  \bibAnnoteFile{NoStop}{sham}%
\bibitem{sham2}%
  \BibitemOpen
  \bibfield{author}{%
  \bibinfo {author} {\bibfnamefont{Y.-H.}\ \bibnamefont{Sham}}, \bibinfo
  {author} {\bibfnamefont{P.~T.}\ \bibnamefont{Leung}},\ and\ \bibinfo {author}
  {\bibfnamefont{L.-M.}\ \bibnamefont{Lin}},\ }%
  \bibfield{journal}{%
  \Doi{10.1103/PhysRevD.87.061503}{\bibinfo {journal} {Phys. Rev. D}}\ }%
  \textbf{\bibinfo {volume} {87}},\ \bibinfo {pages} {061503} (\bibinfo {year}
  {2013})%
  \bibAnnoteFile{NoStop}{sham2}%
\bibitem{structure.exotic.star}%
  \BibitemOpen
  \bibfield{author}{%
  \bibinfo {author} {\bibfnamefont{T.}~\bibnamefont{Harko}}, \bibinfo {author}
  {\bibfnamefont{F.~S.~N.}\ \bibnamefont{Lobo}}, \bibinfo {author}
  {\bibfnamefont{M.~K.}\ \bibnamefont{Mak}},\ and\ \bibinfo {author}
  {\bibfnamefont{S.~V.}\ \bibnamefont{Sushkov}},\ }%
  \bibfield{journal}{%
  \Doi{10.1103/PhysRevD.88.044032}{\bibinfo {journal} {Phys. Rev. D}}\ }%
  \textbf{\bibinfo {volume} {88}},\ \bibinfo {pages} {044032} (\bibinfo {year}
  {2013})%
  \bibAnnoteFile{NoStop}{structure.exotic.star}%
\bibitem{sotani.neutron.star}%
  \BibitemOpen
  \bibfield{author}{%
  \bibinfo {author} {\bibfnamefont{H.}~\bibnamefont{Sotani}},\ }%
  \bibfield{journal}{%
  \Doi{10.1103/PhysRevD.89.104005}{\bibinfo {journal} {Phys. Rev. D}}\ }%
  \textbf{\bibinfo {volume} {89}},\ \bibinfo {pages} {104005} (\bibinfo {year}
  {2014})%
  \bibAnnoteFile{NoStop}{sotani.neutron.star}%
\bibitem{sotani.stellar.oscillations}%
  \BibitemOpen
  \bibfield{author}{%
  \bibinfo {author} {\bibfnamefont{H.}~\bibnamefont{Sotani}},\ }%
  \bibfield{journal}{%
  \Doi{10.1103/PhysRevD.89.124037}{\bibinfo {journal} {Phys. Rev. D}}\ }%
  \textbf{\bibinfo {volume} {89}},\ \bibinfo {pages} {124037} (\bibinfo {year}
  {2014})%
  \bibAnnoteFile{NoStop}{sotani.stellar.oscillations}%
\bibitem{sotani.magnetic.star}%
  \BibitemOpen
  \bibfield{author}{%
  \bibinfo {author} {\bibfnamefont{H.}~\bibnamefont{Sotani}}}%
   (\bibinfo {year} {2015}),\
  \Eprint{http://arxiv.org/abs/1503.07942}{arXiv:1503.07942 [astro-ph.HE]}%
  \bibAnnoteFile{NoStop}{sotani.magnetic.star}%
\bibitem{escamilla}%
  \BibitemOpen
  \bibfield{author}{%
  \bibinfo {author} {\bibfnamefont{C.}~\bibnamefont{Escamilla-Rivera}},
  \bibinfo {author} {\bibfnamefont{M.}~\bibnamefont{Banados}},\ and\ \bibinfo
  {author} {\bibfnamefont{P.~G.}\ \bibnamefont{Ferreira}},\ }%
  \bibfield{journal}{%
  \Doi{10.1103/PhysRevD.85.087302}{\bibinfo {journal} {Phys. Rev. D}}\ }%
  \textbf{\bibinfo {volume} {85}},\ \bibinfo {pages} {087302} (\bibinfo {year}
  {2012})%
  \bibAnnoteFile{NoStop}{escamilla}%
\bibitem{cho}%
  \BibitemOpen
  \bibfield{author}{%
  \bibinfo {author} {\bibfnamefont{I.}~\bibnamefont{Cho}}, \bibinfo {author}
  {\bibfnamefont{H.-C.}\ \bibnamefont{Kim}},\ and\ \bibinfo {author}
  {\bibfnamefont{T.}~\bibnamefont{Moon}},\ }%
  \bibfield{journal}{%
  \Doi{10.1103/PhysRevD.86.084018}{\bibinfo {journal} {Phys. Rev. D}}\ }%
  \textbf{\bibinfo {volume} {86}},\ \bibinfo {pages} {084018} (\bibinfo {year}
  {2012})%
  \bibAnnoteFile{NoStop}{cho}%
\bibitem{avelinoferreira}%
  \BibitemOpen
  \bibfield{author}{%
  \bibinfo {author} {\bibfnamefont{P.~P.}\ \bibnamefont{Avelino}}\ and\
  \bibinfo {author} {\bibfnamefont{R.~Z.}\ \bibnamefont{Ferreira}},\ }%
  \bibfield{journal}{%
  \Doi{10.1103/PhysRevD.86.041501}{\bibinfo {journal} {Phys. Rev. D}}\ }%
  \textbf{\bibinfo {volume} {86}},\ \bibinfo {pages} {041501} (\bibinfo {year}
  {2012})%
  \bibAnnoteFile{NoStop}{avelinoferreira}%
\bibitem{felice}%
  \BibitemOpen
  \bibfield{author}{%
  \bibinfo {author} {\bibfnamefont{A.}~\bibnamefont{De~Felice}}, \bibinfo
  {author} {\bibfnamefont{B.}~\bibnamefont{Gumjudpai}},\ and\ \bibinfo {author}
  {\bibfnamefont{S.}~\bibnamefont{Jhingan}},\ }%
  \bibfield{journal}{%
  \Doi{10.1103/PhysRevD.86.043525}{\bibinfo {journal} {Phys. Rev. D}}\ }%
  \textbf{\bibinfo {volume} {86}},\ \bibinfo {pages} {043525} (\bibinfo {year}
  {2012})%
  \bibAnnoteFile{NoStop}{felice}%
\bibitem{linear.perturbation}%
  \BibitemOpen
  \bibfield{author}{%
  \bibinfo {author} {\bibfnamefont{K.}~\bibnamefont{Yang}}, \bibinfo {author}
  {\bibfnamefont{X.-L.}\ \bibnamefont{Du}},\ and\ \bibinfo {author}
  {\bibfnamefont{Y.-X.}\ \bibnamefont{Liu}},\ }%
  \bibfield{journal}{%
  \Doi{10.1103/PhysRevD.88.124037}{\bibinfo {journal} {Phys. Rev. D}}\ }%
  \textbf{\bibinfo {volume} {88}},\ \bibinfo {pages} {124037} (\bibinfo {year}
  {2013})%
  \bibAnnoteFile{NoStop}{linear.perturbation}%
\bibitem{power.spectrum}%
  \BibitemOpen
  \bibfield{author}{%
  \bibinfo {author} {\bibfnamefont{M.}~\bibnamefont{Lagos}}, \bibinfo {author}
  {\bibfnamefont{M.}~\bibnamefont{Ba\~nados}}, \bibinfo {author}
  {\bibfnamefont{P.~G.}\ \bibnamefont{Ferreira}},\ and\ \bibinfo {author}
  {\bibfnamefont{S.}~\bibnamefont{Garcia-Saenz}},\ }%
  \bibfield{journal}{%
  \Doi{10.1103/PhysRevD.89.024034}{\bibinfo {journal} {Phys. Rev. D}}\ }%
  \textbf{\bibinfo {volume} {89}},\ \bibinfo {pages} {024034} (\bibinfo {year}
  {2014})%
  \bibAnnoteFile{NoStop}{power.spectrum}%
\bibitem{large.scale.structure}%
  \BibitemOpen
  \bibfield{author}{%
  \bibinfo {author} {\bibfnamefont{X.-L.}\ \bibnamefont{Du}}, \bibinfo {author}
  {\bibfnamefont{K.}~\bibnamefont{Yang}}, \bibinfo {author}
  {\bibfnamefont{X.-H.}\ \bibnamefont{Meng}},\ and\ \bibinfo {author}
  {\bibfnamefont{Y.-X.}\ \bibnamefont{Liu}},\ }%
  \bibfield{journal}{%
  \Doi{10.1103/PhysRevD.90.044054}{\bibinfo {journal} {Phys. Rev. D}}\ }%
  \textbf{\bibinfo {volume} {90}},\ \bibinfo {pages} {044054} (\bibinfo {year}
  {2014})%
  \bibAnnoteFile{NoStop}{large.scale.structure}%
\bibitem{bianchi.cosmo}%
  \BibitemOpen
  \bibfield{author}{%
  \bibinfo {author} {\bibfnamefont{T.}~\bibnamefont{Harko}}, \bibinfo {author}
  {\bibfnamefont{F.~S.}\ \bibnamefont{Lobo}},\ and\ \bibinfo {author}
  {\bibfnamefont{M.~K.}\ \bibnamefont{Mak}},\ }%
  \bibfield{journal}{%
  \Doi{10.3390/galaxies2040496}{\bibinfo {journal} {Galaxies}}\ }%
  \textbf{\bibinfo {volume} {2}},\ \bibinfo {pages} {496} (\bibinfo {year}
  {2014})%
  \bibAnnoteFile{NoStop}{bianchi.cosmo}%
\bibitem{eibibrane}%
  \BibitemOpen
  \bibfield{author}{%
  \bibinfo {author} {\bibfnamefont{Y.-X.}\ \bibnamefont{Liu}}, \bibinfo
  {author} {\bibfnamefont{K.}~\bibnamefont{Yang}}, \bibinfo {author}
  {\bibfnamefont{H.}~\bibnamefont{Guo}},\ and\ \bibinfo {author}
  {\bibfnamefont{Y.}~\bibnamefont{Zhong}},\ }%
  \bibfield{journal}{%
  \Doi{10.1103/PhysRevD.85.124053}{\bibinfo {journal} {Phys. Rev. D}}\ }%
  \textbf{\bibinfo {volume} {85}},\ \bibinfo {pages} {124053} (\bibinfo {year}
  {2012})%
  \bibAnnoteFile{NoStop}{eibibrane}%
\bibitem{delsate}%
  \BibitemOpen
  \bibfield{author}{%
  \bibinfo {author} {\bibfnamefont{T.}~\bibnamefont{Delsate}}\ and\ \bibinfo
  {author} {\bibfnamefont{J.}~\bibnamefont{Steinhoff}},\ }%
  \bibfield{journal}{%
  \Doi{10.1103/PhysRevLett.109.021101}{\bibinfo {journal} {Phys. Rev. Lett.}}\
  }%
  \textbf{\bibinfo {volume} {109}},\ \bibinfo {pages} {021101} (\bibinfo {year}
  {2012})%
  \bibAnnoteFile{NoStop}{delsate}%
\bibitem{cho_prd88}%
  \BibitemOpen
  \bibfield{author}{%
  \bibinfo {author} {\bibfnamefont{I.}~\bibnamefont{Cho}}\ and\ \bibinfo
  {author} {\bibfnamefont{H.-C.}\ \bibnamefont{Kim}},\ }%
  \bibfield{journal}{%
  \Doi{10.1103/PhysRevD.88.064038}{\bibinfo {journal} {Phys. Rev. D}}\ }%
  \textbf{\bibinfo {volume} {88}},\ \bibinfo {pages} {064038} (\bibinfo {year}
  {2013})%
  \bibAnnoteFile{NoStop}{cho_prd88}%
\bibitem{jana}%
  \BibitemOpen
  \bibfield{author}{%
  \bibinfo {author} {\bibfnamefont{S.}~\bibnamefont{Jana}}\ and\ \bibinfo
  {author} {\bibfnamefont{S.}~\bibnamefont{Kar}},\ }%
  \bibfield{journal}{%
  \Doi{10.1103/PhysRevD.88.024013}{\bibinfo {journal} {Phys. Rev. D}}\ }%
  \textbf{\bibinfo {volume} {88}},\ \bibinfo {pages} {024013} (\bibinfo {year}
  {2013})%
  \bibAnnoteFile{NoStop}{jana}%
\bibitem{solar.test}%
  \BibitemOpen
  \bibfield{author}{%
  \bibinfo {author} {\bibfnamefont{J.}~\bibnamefont{Casanellas}}, \bibinfo
  {author} {\bibfnamefont{P.}~\bibnamefont{Pani}}, \bibinfo {author}
  {\bibfnamefont{I.}~\bibnamefont{Lopes}},\ and\ \bibinfo {author}
  {\bibfnamefont{V.}~\bibnamefont{Cardoso}},\ }%
  \bibfield{journal}{%
  \bibinfo {journal} {The Astrophysical Journal}\ }%
  \textbf{\bibinfo {volume} {745}},\ \bibinfo {pages} {15} (\bibinfo {year}
  {2012})%
  \bibAnnoteFile{NoStop}{solar.test}%
\bibitem{nuclear.test}%
  \BibitemOpen
  \bibfield{author}{%
  \bibinfo {author} {\bibfnamefont{P.}~\bibnamefont{Avelino}},\ }%
  \bibfield{journal}{%
  \bibinfo {journal} {Journal of Cosmology and Astroparticle Physics}\ }%
  \textbf{\bibinfo {volume} {2012}},\ \bibinfo {pages} {022} (\bibinfo {year}
  {2012})%
  \bibAnnoteFile{NoStop}{nuclear.test}%
\bibitem{pani}%
  \BibitemOpen
  \bibfield{author}{%
  \bibinfo {author} {\bibfnamefont{P.}~\bibnamefont{Pani}}\ and\ \bibinfo
  {author} {\bibfnamefont{T.~P.}\ \bibnamefont{Sotiriou}},\ }%
  \bibfield{journal}{%
  \Doi{10.1103/PhysRevLett.109.251102}{\bibinfo {journal} {Phys. Rev. Lett.}}\
  }%
  \textbf{\bibinfo {volume} {109}},\ \bibinfo {pages} {251102} (\bibinfo {year}
  {2012})%
  \bibAnnoteFile{NoStop}{pani}%
\bibitem{eibiprob.cure}%
  \BibitemOpen
  \bibfield{author}{%
  \bibinfo {author} {\bibfnamefont{H.-C.}\ \bibnamefont{Kim}},\ }%
  \bibfield{journal}{%
  \Doi{10.1103/PhysRevD.89.064001}{\bibinfo {journal} {Phys. Rev. D}}\ }%
  \textbf{\bibinfo {volume} {89}},\ \bibinfo {pages} {064001} (\bibinfo {year}
  {2014})%
  \bibAnnoteFile{NoStop}{eibiprob.cure}%
\bibitem{odintsov}%
  \BibitemOpen
  \bibfield{author}{%
  \bibinfo {author} {\bibfnamefont{S.~D.}\ \bibnamefont{Odintsov}}, \bibinfo
  {author} {\bibfnamefont{G.~J.}\ \bibnamefont{Olmo}},\ and\ \bibinfo {author}
  {\bibfnamefont{D.}~\bibnamefont{Rubiera-Garcia}},\ }%
  \bibfield{journal}{%
  \Doi{10.1103/PhysRevD.90.044003}{\bibinfo {journal} {Phys. Rev. D}}\ }%
  \textbf{\bibinfo {volume} {90}},\ \bibinfo {pages} {044003} (\bibinfo {year}
  {2014})%
  \bibAnnoteFile{NoStop}{odintsov}%
\bibitem{fernandes}%
  \BibitemOpen
  \bibfield{author}{%
  \bibinfo {author} {\bibfnamefont{K.}~\bibnamefont{Fernandes}}\ and\ \bibinfo
  {author} {\bibfnamefont{A.}~\bibnamefont{Lahiri}},\ }%
  \bibfield{journal}{%
  \Doi{10.1103/PhysRevD.91.044014}{\bibinfo {journal} {Phys. Rev. D}}\ }%
  \textbf{\bibinfo {volume} {91}},\ \bibinfo {pages} {044014} (\bibinfo {year}
  {2015})%
  \bibAnnoteFile{NoStop}{fernandes}%
\bibitem{wei}%
  \BibitemOpen
  \bibfield{author}{%
  \bibinfo {author} {\bibfnamefont{S.-W.}\ \bibnamefont{Wei}}, \bibinfo
  {author} {\bibfnamefont{K.}~\bibnamefont{Yang}},\ and\ \bibinfo {author}
  {\bibfnamefont{Y.-X.}\ \bibnamefont{Liu}}}%
   (\bibinfo {year} {2014}),\
  \Eprint{http://arxiv.org/abs/1405.2178}{arXiv:1405.2178 [gr-qc]}%
  \bibAnnoteFile{NoStop}{wei}%
\bibitem{sotani}%
  \BibitemOpen
  \bibfield{author}{%
  \bibinfo {author} {\bibfnamefont{H.}~\bibnamefont{Sotani}}\ and\ \bibinfo
  {author} {\bibfnamefont{U.}~\bibnamefont{Miyamoto}},\ }%
  \bibfield{journal}{%
  \Doi{10.1103/PhysRevD.90.124087}{\bibinfo {journal} {Phys. Rev. D}}\ }%
  \textbf{\bibinfo {volume} {90}},\ \bibinfo {pages} {124087} (\bibinfo {year}
  {2014})%
  \bibAnnoteFile{NoStop}{sotani}%
\bibitem{eibiwormhole}%
  \BibitemOpen
  \bibfield{author}{%
  \bibinfo {author} {\bibfnamefont{G.~J.}\ \bibnamefont{Olmo}}, \bibinfo
  {author} {\bibfnamefont{D.}~\bibnamefont{Rubiera-Garcia}},\ and\ \bibinfo
  {author} {\bibfnamefont{H.}~\bibnamefont{Sanchis-Alepuz}},\ }%
  \bibfield{journal}{%
  \Doi{10.1140/epjc/s10052-014-2804-8}{\bibinfo {journal} {The European
  Physical Journal C}}\ }%
  \textbf{\bibinfo {volume} {74}},\ \bibinfo {pages} {2804} (\bibinfo {year}
  {2014})%
  \bibAnnoteFile{NoStop}{eibiwormhole}%
\bibitem{rajibul}%
  \BibitemOpen
  \bibfield{author}{%
  \bibinfo {author} {\bibfnamefont{R.}~\bibnamefont{Shaikh}},\ }%
  \bibfield{journal}{%
  \Doi{10.1103/PhysRevD.92.024015}{\bibinfo {journal} {Phys. Rev. D}}\ }%
  \textbf{\bibinfo {volume} {92}},\ \bibinfo {pages} {024015} (\bibinfo {year}
  {2015})%
  \bibAnnoteFile{NoStop}{rajibul}%
\bibitem{geon}%
  \BibitemOpen
  \bibfield{author}{%
  \bibinfo {author} {\bibfnamefont{M.}~\bibnamefont{Demianski}},\ }%
  \bibfield{journal}{%
  \Doi{10.1007/BF01889380}{\bibinfo {journal} {Foundations of Physics}}\ }%
  \textbf{\bibinfo {volume} {16}},\ \bibinfo {pages} {187} (\bibinfo {year}
  {1986})%
  \bibAnnoteFile{NoStop}{geon}%
\bibitem{Einstein.BI.1}%
  \BibitemOpen
  \bibfield{author}{%
  \bibinfo {author} {\bibfnamefont{T.}~\bibnamefont{Tamaki}}\ and\ \bibinfo
  {author} {\bibfnamefont{T.}~\bibnamefont{Torii}},\ }%
  \bibfield{journal}{%
  \Doi{10.1103/PhysRevD.62.061501}{\bibinfo {journal} {Phys. Rev. D}}\ }%
  \textbf{\bibinfo {volume} {62}},\ \bibinfo {pages} {061501} (\bibinfo {year}
  {2000})%
  \bibAnnoteFile{NoStop}{Einstein.BI.1}%
\bibitem{Einstein.BI.2}%
  \BibitemOpen
  \bibfield{author}{%
  \bibinfo {author} {\bibfnamefont{R.}~\bibnamefont{Yamazaki}}\ and\ \bibinfo
  {author} {\bibfnamefont{D.}~\bibnamefont{Ida}},\ }%
  \bibfield{journal}{%
  \Doi{10.1103/PhysRevD.64.024009}{\bibinfo {journal} {Phys. Rev. D}}\ }%
  \textbf{\bibinfo {volume} {64}},\ \bibinfo {pages} {024009} (\bibinfo {year}
  {2001})%
  \bibAnnoteFile{NoStop}{Einstein.BI.2}%
\bibitem{Einstein.BI.3}%
  \BibitemOpen
  \bibfield{author}{%
  \bibinfo {author} {\bibfnamefont{S.~H.}\ \bibnamefont{Mazharimousavi}}\ and\
  \bibinfo {author} {\bibfnamefont{M.}~\bibnamefont{Halilsoy}}}%
   (\bibinfo {year} {2014}),\
  \Eprint{http://arxiv.org/abs/1405.2956}{arXiv:1405.2956 [gr-qc]}%
  \bibAnnoteFile{NoStop}{Einstein.BI.3}%
\bibitem{dibakar1}%
  \BibitemOpen
  \bibfield{author}{%
  \bibinfo {author} {\bibfnamefont{R.}~\bibnamefont{Banerjee}}\ and\ \bibinfo
  {author} {\bibfnamefont{D.}~\bibnamefont{Roychowdhury}},\ }%
  \bibfield{journal}{%
  \Doi{10.1103/PhysRevD.85.104043}{\bibinfo {journal} {Phys. Rev. D}}\ }%
  \textbf{\bibinfo {volume} {85}},\ \bibinfo {pages} {104043} (\bibinfo {year}
  {2012})%
  \bibAnnoteFile{NoStop}{dibakar1}%
\bibitem{dibakar2}%
  \BibitemOpen
  \bibfield{author}{%
  \bibinfo {author} {\bibfnamefont{A.}~\bibnamefont{Lala}}\ and\ \bibinfo
  {author} {\bibfnamefont{D.}~\bibnamefont{Roychowdhury}},\ }%
  \bibfield{journal}{%
  \Doi{10.1103/PhysRevD.86.084027}{\bibinfo {journal} {Phys. Rev. D}}\ }%
  \textbf{\bibinfo {volume} {86}},\ \bibinfo {pages} {084027} (\bibinfo {year}
  {2012})%
  \bibAnnoteFile{NoStop}{dibakar2}%
\bibitem{dibakar3}%
  \BibitemOpen
  \bibfield{author}{%
  \bibinfo {author} {\bibfnamefont{R.}~\bibnamefont{Banerjee}}\ and\ \bibinfo
  {author} {\bibfnamefont{D.}~\bibnamefont{Roychowdhury}},\ }%
  \bibfield{journal}{%
  \Doi{10.1103/PhysRevD.85.044040}{\bibinfo {journal} {Phys. Rev. D}}\ }%
  \textbf{\bibinfo {volume} {85}},\ \bibinfo {pages} {044040} (\bibinfo {year}
  {2012})%
  \bibAnnoteFile{NoStop}{dibakar3}%
\bibitem{jonas}%
  \BibitemOpen
  \bibfield{author}{%
  \bibinfo {author} {\bibfnamefont{J.~P.}\ \bibnamefont{Pereira}}\ and\
  \bibinfo {author} {\bibfnamefont{J.~A.}\ \bibnamefont{Rueda}},\ }%
  \bibfield{journal}{%
  \Doi{10.1103/PhysRevD.91.064048}{\bibinfo {journal} {Phys. Rev. D}}\ }%
  \textbf{\bibinfo {volume} {91}},\ \bibinfo {pages} {064048} (\bibinfo {year}
  {2015})%
  \bibAnnoteFile{NoStop}{jonas}%
\bibitem{openstring1}%
  \BibitemOpen
  \bibfield{author}{%
  \bibinfo {author} {\bibfnamefont{E.}~\bibnamefont{Fradkin}}\ and\ \bibinfo
  {author} {\bibfnamefont{A.}~\bibnamefont{Tseytlin}},\ }%
  \bibfield{journal}{%
  \Doi{http://dx.doi.org/10.1016/0370-2693(85)90205-9}{\bibinfo {journal}
  {Physics Letters B}}\ }%
  \textbf{\bibinfo {volume} {163}},\ \bibinfo {pages} {123 } (\bibinfo {year}
  {1985})%
  \bibAnnoteFile{NoStop}{openstring1}%
\bibitem{openstring2}%
  \BibitemOpen
  \bibfield{author}{%
  \bibinfo {author} {\bibfnamefont{A.}~\bibnamefont{Abouelsaood}}, \bibinfo
  {author} {\bibfnamefont{C.}~\bibnamefont{Callan}}, \bibinfo {author}
  {\bibfnamefont{C.}~\bibnamefont{Nappi}},\ and\ \bibinfo {author}
  {\bibfnamefont{S.}~\bibnamefont{Yost}},\ }%
  \bibfield{journal}{%
  \Doi{http://dx.doi.org/10.1016/0550-3213(87)90164-7}{\bibinfo {journal}
  {Nuclear Physics B}}\ }%
  \textbf{\bibinfo {volume} {280}},\ \bibinfo {pages} {599 } (\bibinfo {year}
  {1987})%
  \bibAnnoteFile{NoStop}{openstring2}%
\bibitem{wald}%
  \BibitemOpen
  \bibfield{author}{%
  \bibinfo {author} {\bibfnamefont{R.~M.}\ \bibnamefont{Wald}},\ }%
  \emph{\bibinfo {title} {General Relativity}}\ (\bibinfo {publisher}
  {University of Chicago Press},\ \bibinfo {address} {Chacago and London},\
  \bibinfo {year} {1984})%
  \bibAnnoteFile{NoStop}{wald}%
\bibitem{plebansky}%
  \BibitemOpen
  \bibfield{author}{%
  \bibinfo {author} {\bibfnamefont{J.~F.}\ \bibnamefont{Plebansky}},\ }%
  \emph{\bibinfo {title} {Lectures on Nonlinear Electrodynamics}}\ (\bibinfo
  {address} {Nordita, Copenhagen},\ \bibinfo {year} {1968})%
  \bibAnnoteFile{NoStop}{plebansky}%
\bibitem{novello}%
  \BibitemOpen
  \bibfield{author}{%
  \bibinfo {author} {\bibfnamefont{M.}~\bibnamefont{Novello}}, \bibinfo
  {author} {\bibfnamefont{V.~A.}\ \bibnamefont{De~Lorenci}}, \bibinfo {author}
  {\bibfnamefont{J.~M.}\ \bibnamefont{Salim}},\ and\ \bibinfo {author}
  {\bibfnamefont{R.}~\bibnamefont{Klippert}},\ }%
  \bibfield{journal}{%
  \Doi{10.1103/PhysRevD.61.045001}{\bibinfo {journal} {Phys. Rev. D}}\ }%
  \textbf{\bibinfo {volume} {61}},\ \bibinfo {pages} {045001} (\bibinfo {year}
  {2000})%
  \bibAnnoteFile{NoStop}{novello}%
\bibitem{virbhadra3}%
  \BibitemOpen
  \bibfield{author}{%
  \bibinfo {author} {\bibfnamefont{C.-M.}\ \bibnamefont{Claudel}}, \bibinfo
  {author} {\bibfnamefont{K.~S.}\ \bibnamefont{Virbhadra}},\ and\ \bibinfo
  {author} {\bibfnamefont{G.~F.~R.}\ \bibnamefont{Ellis}},\ }%
  \bibfield{journal}{%
  \Doi{10.1063/1.1308507}{\bibinfo {journal} {JOURNAL OF MATHEMATICAL
  PHYSICS}}\ }%
  \textbf{\bibinfo {volume} {42}},\ \bibinfo {pages} {2} (\bibinfo {year}
  {2001})%
  \bibAnnoteFile{NoStop}{virbhadra3}%
\bibitem{virbhadra4}%
  \BibitemOpen
  \bibfield{author}{%
  \bibinfo {author} {\bibfnamefont{K.~S.}\ \bibnamefont{Virbhadra}}\ and\
  \bibinfo {author} {\bibfnamefont{G.~F.~R.}\ \bibnamefont{Ellis}},\ }%
  \bibfield{journal}{%
  \Doi{10.1103/PhysRevD.65.103004}{\bibinfo {journal} {Phys. Rev. D}}\ }%
  \textbf{\bibinfo {volume} {65}},\ \bibinfo {pages} {103004} (\bibinfo {year}
  {2002})%
  \bibAnnoteFile{NoStop}{virbhadra4}%
\bibitem{virbhadra1}%
  \BibitemOpen
  \bibfield{author}{%
  \bibinfo {author} {\bibfnamefont{K.~S.}\ \bibnamefont{Virbhadra}},\ }%
  \bibfield{journal}{%
  \Doi{10.1103/PhysRevD.79.083004}{\bibinfo {journal} {Phys. Rev. D}}\ }%
  \textbf{\bibinfo {volume} {79}},\ \bibinfo {pages} {083004} (\bibinfo {year}
  {2009})%
  \bibAnnoteFile{NoStop}{virbhadra1}%
\bibitem{virbhadra2}%
  \BibitemOpen
  \bibfield{author}{%
  \bibinfo {author} {\bibfnamefont{K.~S.}\ \bibnamefont{Virbhadra}}\ and\
  \bibinfo {author} {\bibfnamefont{G.~F.~R.}\ \bibnamefont{Ellis}},\ }%
  \bibfield{journal}{%
  \bibinfo {journal} {Phys. Rev. D}\ }%
  \textbf{\bibinfo {volume} {62}},\ \bibinfo {pages} {084003} (\bibinfo {year}
  {2000})%
  \bibAnnoteFile{NoStop}{virbhadra2}%
\bibitem{bodenner}%
  \BibitemOpen
  \bibfield{author}{%
  \bibinfo {author} {\bibfnamefont{J.}~\bibnamefont{Bodenner}}\ and\ \bibinfo
  {author} {\bibfnamefont{C.~M.}\ \bibnamefont{Will}},\ }%
  \bibfield{journal}{%
  \Doi{http://dx.doi.org/10.1119/1.1570416}{\bibinfo {journal} {American
  Journal of Physics}}\ }%
  \textbf{\bibinfo {volume} {71}},\ \bibinfo {pages} {770} (\bibinfo {year}
  {2003})%
  \bibAnnoteFile{NoStop}{bodenner}%
\bibitem{briet}%
  \BibitemOpen
  \bibfield{author}{%
  \bibinfo {author} {\bibfnamefont{J.}~\bibnamefont{Briet}}\ and\ \bibinfo
  {author} {\bibfnamefont{D.}~\bibnamefont{Hobill}}}%
   (\bibinfo {year} {2008}),\
  \Eprint{http://arxiv.org/abs/0801.3859}{arXiv:0801.3859 [gr-qc]}%
  \bibAnnoteFile{NoStop}{briet}%
\bibitem{bozza}%
  \BibitemOpen
  \bibfield{author}{%
  \bibinfo {author} {\bibfnamefont{V.}~\bibnamefont{Bozza}}, \bibinfo {author}
  {\bibfnamefont{S.}~\bibnamefont{Capozziello}}, \bibinfo {author}
  {\bibfnamefont{G.}~\bibnamefont{Iovane}},\ and\ \bibinfo {author}
  {\bibfnamefont{G.}~\bibnamefont{Scarpetta}},\ }%
  \bibfield{journal}{%
  \Doi{10.1023/A:1012292927358}{\bibinfo {journal} {General Relativity and
  Gravitation}}\ }%
  \textbf{\bibinfo {volume} {33}},\ \bibinfo {pages} {1535} (\bibinfo {year}
  {2001})%
  \bibAnnoteFile{NoStop}{bozza}%
\end{thebibliography}%


%
\end{document}